%% file: main.tex
\documentclass[sigconf]{acmart}
\usepackage{booktabs}       % professional-quality tables
\usepackage{mathtools}
\usepackage{xspace}
\usepackage{multicol}
\usepackage{multirow}
\usepackage{algorithm}
\usepackage[noend]{algpseudocode}
\usepackage{enumitem}
\input{TEX/dfn}

\AtBeginDocument{%
  \providecommand\BibTeX{{%
    \normalfont B\kern-0.5em{\scshape i\kern-0.25em b}\kern-0.8em\TeX}}}

\setcopyright{acmcopyright}
\copyrightyear{2022}
\acmYear{2022}
\acmDOI{10.1145/1122445.1122456}

\acmConference[xxxx 'xx]{ACM xxx 0000}{xxx 00 -- 99, 0000}{xx, xx}
\acmBooktitle{xxx}
\acmPrice{15.00}
\acmISBN{978-1-4503-XXXX-X/22/04}

\begin{document}

%%
%% The "title" command has an optional parameter,
%% allowing the author to define a "short title" to be used in page headers.
%\title{Not All Documents are Created Equal: A Generative Model and A Meta Algorithm for Dense Retrieval}
% Behavioral Representations helps Memorization for Long-tailed Dense Retrieval
%Dense Retrieval with Non-Uniform Distribution of Representational Budget
\title{On the Value of Behavioral Representations for Dense Retrieval}

%%
%% The "author" command and its associated commands are used to define
%% the authors and their affiliations.
%% Of note is the shared affiliation of the first two authors, and the
%% "authornote" and "authornotemark" commands
%% used to denote shared contribution to the research.
\author{Nan Jiang}
\authornote{Both authors contributed equally to the paper}
\affiliation{%
 \institution{Purdue University}
 \city{}
 \state{IN}
 \country{USA}
 \postcode{}
}
\email{jiang631@purdue.edu}

\author{Dhivya Eswaran}
\authornotemark[1]
\affiliation{%
 \institution{Amazon Search}
 \city{}
 \state{CA}
 \country{USA}
 \postcode{}
}
\email{deswaran@amazon.com}

\author{Choon Hui Teo}
\affiliation{%
 \institution{Amazon Search}
 \city{}
 \state{CA}
 \country{USA}
 \postcode{}
}
\email{choonhui@amazon.com}

\author{Yexiang Xue}
\affiliation{%
 \institution{Purdue University}
 \city{}
 \state{IN}
 \country{USA}
 \postcode{}
}
\email{yexiang@purdue.edu}

\author{Yesh Dattatreya}
\affiliation{%
 \institution{Amazon Search}
 \city{}
 \state{CA}
 \country{USA}
 \postcode{}
}
\email{ydatta@amazon.com}

\author{Sujay Sanghav}
\affiliation{%
 \institution{Amazon Search}
 \city{}
 \state{CA}
 \country{USA}
 \postcode{}
}
\email{sujayrs@amazon.com}

\author{Vishy Vishwanathan}
\affiliation{%
 \institution{Amazon Search}
 \city{}
 \state{CA}
 \country{USA}
 \postcode{}
}
\email{vishy@amazon.com}

% \begin{CCSXML}
% <ccs2012>
% <concept>
% <concept_id>10002951.10003317</concept_id>
% <concept_desc>Information systems~Information retrieval</concept_desc>
% <concept_significance>500</concept_significance>
% </concept>
% <concept>
% <concept_id>10002951.10003317.10003318</concept_id>
% <concept_desc>Information systems~Document representation</concept_desc>
% <concept_significance>500</concept_significance>
% </concept>
% </ccs2012>
% \end{CCSXML}

% \ccsdesc[500]{Information systems~Information retrieval}
% \ccsdesc[500]{Information systems~Document representation}

%%
%% By default, the full list of authors will be used in the page
%% headers. Often, this list is too long, and will overlap
%% other information printed in the page headers. This command allows
%% the author to define a more concise list
%% of authors' names for this purpose.
\renewcommand{\shortauthors}{}

%%
%% The abstract is a short summary of the work to be presented in the
%% article.
\begin{abstract}

\input{TEX/0-abstract}

\end{abstract}

%%
%% The code below is generated by the tool at http://dl.acm.org/ccs.cfm.
%% Please copy and paste the code instead of the example below.
%%

%%
%% Keywords. The author(s) should pick words that accurately describe
%% the work being presented. Separate the keywords with commas.

% \keywords{Dense Retrieval; Multi-view Representation; Heavy-tailed Distribution}

%%
%% This command processes the author and affiliation and title
%% information and builds the first part of the formatted document.
\maketitle

\section{Introduction}
\label{sec:introduction}
\input{TEX/1-introduction}

\section{Related Work}
\label{sec:relatedwork}
\input{TEX/2-relatedwork}

\input{TEX/3-prelimnaires}

\input{TEX/4-generative}

\input{TEX/5-method}

\section{Experiments}
\label{sec:experiments}
\input{TEX/6-experiments}

\section{Conclusion and Future Work}
\label{sec:conclusion}

\input{TEX/8-conclusion}

%%
%% If your work has an appendix, this is the place to put it.

\bibliographystyle{ACM-Reference-Format}
\bibliography{ref}

\end{document}

%% file: TEX/dfn.tex
\usepackage{pifont}% http://ctan.org/pkg/pifont
\newcommand{\cmark}{\ding{51}}%
\newcommand{\xmark}{\ding{55}}%

\newcommand{\method}{\textsc{MVG}\xspace}
\newcommand{\mvg}{\textsc{MVG}\xspace}
\newcommand{\mebert}{\textsc{ME-BERT}\xspace}
\newcommand{\colbert}{ColBERT\xspace}

\newcommand{\methodfull}{Multi-View Geometric Index\xspace}
\newcommand{\mywidth}{}

\newcommand{\refsec}[1]{Section~\ref{sec:#1}}
\newcommand{\reffig}[1]{Figure~\ref{fig:#1}}
\newcommand{\reflem}[1]{Lemma~\ref{lem:#1}}
\newcommand{\reftab}[1]{Table~\ref{tab:#1}}

\newcommand{\refalg}[1]{Algorithm~\ref{alg:#1}}
\renewcommand{\refeq}[1]{Equation~\eqref{eq:#1}}
\newtheorem{problem}{Problem}
\newtheorem{lemma}{Lemma}
\newtheorem{corollary}{Corollary}

\newcommand{\dssm}{\textsc{DSSM}\xspace}

\newcommand{\sbert}{\textsc{S.Bert}\xspace}
\newcommand{\prodmodel}{{\textsc{Prod}}\xspace}

\newcommand{\BertBase}{{\textsc{BertBase}}\xspace}
\newcommand{\LegalBert}{{\textsc{LegalBert}}\xspace}

\newcommand{\queryset}{\mathcal{Q}\xspace}
\newcommand{\docset}{\mathcal{D}\xspace}

\newcommand{\weight}{w\xspace}
\newcommand{\query}{q\xspace}
\newcommand{\doc}{d\xspace}

\newcommand{\modelvec}[1]{\mathbf{x}_{#1}\xspace}

\newcommand{\dimension}{r\space}

\newcommand{\budget}{M \xspace}

\newcommand{\cidvec}[1]{\mathbf{v}_{#1}\xspace}

\newcommand{\numcidvec}{m}

\newcommand{\eurlex}{EurLex\xspace}
\newcommand{\wikiseealso}{WikiSeeAlso\xspace}
\newcommand{\pubmed}{PubMed\xspace}

\newcommand{\prodsearch}{ProdSearch\xspace}

%% file: TEX/0-abstract.tex
We consider text retrieval within dense representational space in real-world settings such as e-commerce search where (a) document popularity and (b) diversity of queries associated with a document have a skewed distribution. Most of contemporary dense retrieval literature--which typically focuses on MSMARCO and TREC benchmark datasets--present two shortcomings in these settings. (1)~They learn an \textit{almost equal number of representations per document}, agnostic to the fact that a few `head' documents are disproportionately more critical to achieving a good retrieval performance. (ii)~They learn \textit{purely semantic document representations} inferred from intrinsic document characteristics (e.g. tokenized text) which may not contain adequate information to determine the queries for which the document is relevant--especially when the document is short.

We propose to overcome these limitations by \textit{augmenting semantic document representations learned by bi-encoders with behavioral document representations} learned by our proposed approach \method. To do so, \method (1) determines how to divide the total budget for behavioral representations by drawing a connection to Pitman-Yor process, and (2) simply clusters the queries related to a given document (based on user behavior) within the representational space learned by a base bi-encoder, and treats the cluster centers as its behavioral representations. Our central contribution is the finding \textit{such a simple intuitive light-weight approach leads to substantial gains in key first-stage retrieval metrics (e.g. recall) by incurring only a marginal memory overhead}. We establish this via extensive experiments over three large public datasets comparing to several single-vector (e.g. SentenceBERT) and multi-vector (e.g. ColBERT) bi-encoders, a proprietary e-commerce search dataset comparing to production-quality bi-encoder, and an A/B test. We hope that the next generation of dense retrieval approaches more carefully considers the dual questions of representational budget distribution, and of jointly learning semantic and behavioral representations.

%% file: TEX/1-introduction.tex
Dense retrieval (DR) is a powerful workhorse for large-scale language systems which rely on text retrieval as the first step, e.g. web search~\cite{DBLP:conf/iclr/XiongXLTLBAO21,DBLP:journals/tacl/LuanETC21}, e-commerce shopping~\cite{DBLP:conf/kdd/NigamSMLDSTGY19} and multi-label text categorization~\cite{DBLP:conf/wsdm/MittalDASAKV21}. The main idea in DR is to embed queries and documents into a common continuous representation space, such that the relevance of a document to a query is captured by the proximity of their representations in this space. The representations are robust and fully learnable from data; thus DR greatly enhances \textbf{generalization} (e.g. robustness in the face of synonyms, misspellings, morphological variants) compared to sparse retrieval approaches which rely on lexical overlap, e.g. BM25~\cite{DBLP:journals/jasis/RobertsonJ76} . During serving, the document representations are pre-computed and stored within an efficient nearest-neighbor (NN) index for fast lookup~\cite{DBLP:journals/tbd/JohnsonDJ21}.

\begin{figure}[!tb]
	\centering
	\renewcommand{\mywidth}{0.495\columnwidth}
	\resizebox{1\columnwidth}{!}{%
		\begin{tabular}{ccc}
			\includegraphics[height=\mywidth]{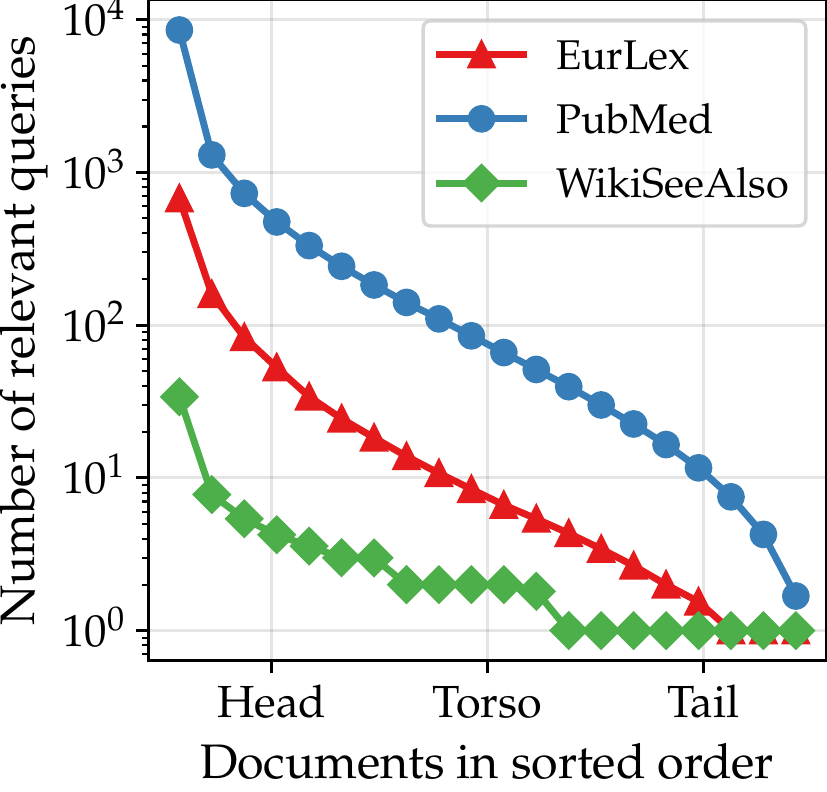} &
			\includegraphics[height=\mywidth]{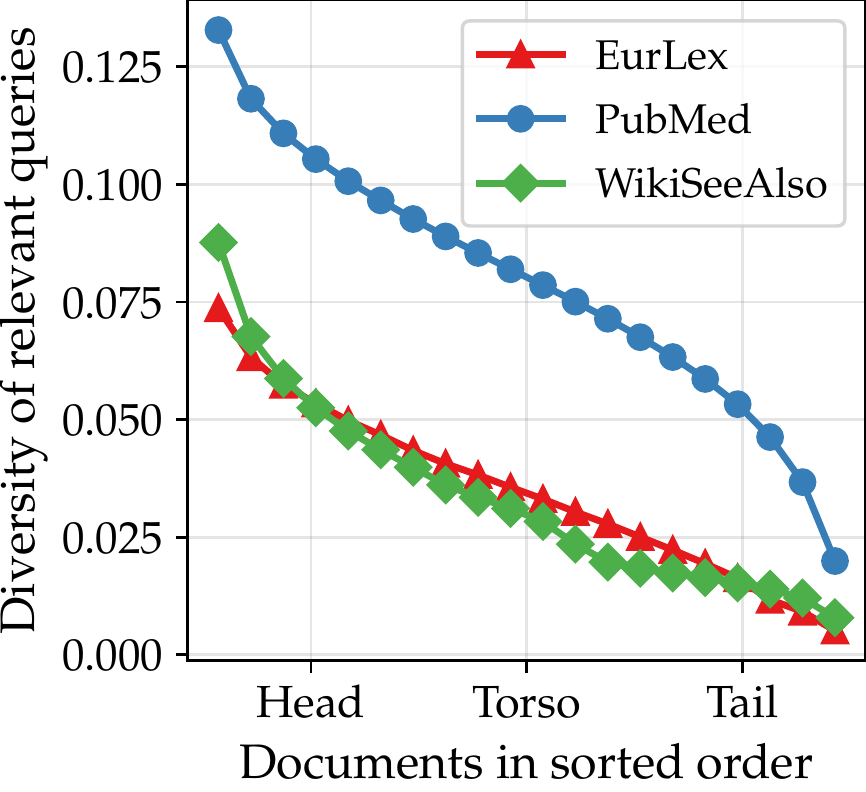}
		\end{tabular}%
	}
	\caption{Document popularity distributions in a few common datasets: A few documents are seen to be disproportionately \textit{popular} (left) and relevant to a \textit{diverse} (right) set of queries. This motivates us to consider non-uniform number of extra (behavioral) representations for such documents within the nearest-neighbor (NN) index of dense retrieval.}
	\label{fig:motivation}
	\vspace{-4mm}
\end{figure}

Our motivating observation is that \textit{in many real-world retrieval settings, document popularity follows a skewed (e.g. power-law) distribution
~\cite{DBLP:conf/sigcomm/FaloutsosFF99}}. As shown in \reffig{motivation} (left), most user queries result in clicks for only a small set of `head' documents whereas there is a long `tail' of documents with very few associated queries. Thus \textbf{memorization} of associations between these \textit{head} documents and queries associated with them in the past (as well as generalizing these associations to new query variants) tends to be disproportionately more important to maintaining a good retrieval performance in real-world settings, compared to the retrieval of \textit{tail} documents. 

Typical DR approaches~\cite{DBLP:conf/cikm/HuangHGDAH13,DBLP:journals/corr/MitraNCC16,DBLP:conf/emnlp/ReimersG19,DBLP:conf/kdd/NigamSMLDSTGY19,DBLP:conf/kdd/HaldarRSAZMYTL20,DBLP:conf/iclr/XiongXLTLBAO21} are `single-vector' i.e. they learn a single representation per document, agnostic to whether the document is head or tail. However, \textit{learning only a single representation per head document is not adequate for memorization of its historical query associations.} \reffig{motivation} (right) compares the diversity of queries used to access head and tail documents, and shows that head documents are typically accessed through a disproportionately \textit{diverse} set of queries compared to tail documents\footnote{Here, diversity of a set of queries is measured by the volume of the smallest bounding box enclosing all their representations in a given space (we use \texttt{bert-base-uncased} model from \url{https://huggingface.co}). Concretely, if $l_j$ is the length of this box along dimension $j \in \{1,\ldots,r\}$, we compute diversity as $(l_1 l_2 \ldots l_\dimension)^{1/\dimension}$. But other notions of diversity (e.g. smallest bounding ball instead of box) led to a similar observation.}. 
However, such diverse and disparate reasons for accessing a head document cannot be shoehorned into a single representation: triangle inequality (a typical prerequisite for the representation space to be well-defined) disallows a document to be close to two queries, when the queries are themselves well-separated in this space.

Recent DR approaches~\cite{DBLP:conf/acl/TangSJWZW20,DBLP:journals/tacl/LuanETC21,DBLP:conf/sigir/KhattabZ20,DBLP:journals/naacl/maize} consider learning multiple representations per document. Still, \textit{a prevailing limitation is that the learned document representations are purely semantic}. That is, once the `behavioral' graph of query-document associations is used to train the bi-encoder, the graph is discarded, and only the documents' intrinsic features (typically, tokenized text) are used to infer their representations. This is especially problematic for short documents where the tokenized text does not contain adequate information to determine the queries for which it is relevant. Graph neural networks~\cite{DBLP:conf/kdd/LiuTLYZH19,DBLP:conf/kdd/YangPZP0RL20,DBLP:conf/www/FanLL00LLLJY22} is an attractive framework to better leverage graph information, but suffers from high online retrieval latency~\cite{DBLP:journals/corr/2022graph}.

To address the above challenges, we propose to \textit{augment semantic document representations learned by bi-encoder models with behavioral document representations learned by our proposed approach: \method (Multi-View Geometric Index)}.
\method and its resulting behavioral representations have several attractive properties: \textit{\textbf{(a)~Variable number of representations per document:}} The number of behavioral representations learned per document can be controlled via a hyperparameter to be as skewed as is appropriate for a given dataset. \textbf{\textit{(b)~Improved memorization for head documents:}} The behavioral representations are learned directly from the behavioral graph (hence the name) by simply clustering the queries connected to the document within the representational space given by the bi-encoder. Thus, behavioral representations are not tied to intrinsic document features, and are more flexible in memorizing query-document associations. \textbf{\textit{(c)~Light-weight practical approach:}} \mvg can be implemented with a light-weight step (step 3) in the standard DR pipeline as shown in \reftab{dense_retrieval_pipeline}, with no change in bi-encoder training or inference procedure (steps 1-2), the NN-index building step (step 4), or the online retrieval logic (step 5). These properties, together, set \method apart from past dense retrieval literature as we illustrate in \reftab{relatedwork} and detail in \refsec{relatedwork}.

\begin{table}[!tb]
    \centering
    \caption{\method-based Dense Retrieval with steps 1-4 performed offline and step 5 conducted online. Only step 3 (focus of this paper) is new compared to standard dense retrieval pipeline.\vspace{-2mm}}
    \label{tab:dense_retrieval_pipeline}
    \begin{tabular}{p{7cm}|c}
        \hline%\toprule
        \textbf{Steps in \method-based Dense Retrieval} & \textbf{New?}  \\
        \hline %\midrule 
        1. Learn a bi-encoder model $\mathcal{M}$ & \xmark\\
        2. Infer semantic document representations from $\mathcal{M}$ & \xmark\\
        3. Infer behavioral document representations from \method & \cmark\\
        4. Build an NN-index of all document representations & \xmark\\
        5. Query the NN-index online & \xmark\\
        \hline %\bottomrule
    \end{tabular}
    \vspace{-3mm}
\end{table}

Concretely, our contributions are four-fold:

\begin{itemize}[align=left, leftmargin=0pt, labelwidth=0pt, itemindent=!]
    \item 
    \textit{\textbf{Theoretical Connection:}} We leverage Pitman-Yor Process~\cite{10.1214/aop/1024404422} (PYP) to explain the skewed distribution of query diversity from \reffig{motivation}. Under PYP, the expected number of query clusters $m$ for a document with $n$ queries scales as $m = \mathcal{O}(n^{\beta})$ for some $\beta\in(0,1)$. 

    \item 
    \textit{\textbf{Light-Weight Algorithm:}} We propose the \method algorithm to improve the recall of \textit{any} bi-encoder by incurring a marginal overhead in NN-index size. The core algorithmic questions here are: (i)~how many behavioral representations to learn per document, and (ii)~how to learn them. \method provides principled answers for both, by leveraging the connection to PYP for (i) and formulating (ii) as a constrained clustering problem on a unit sphere. \method is also easy to deploy within any industrial infrastructure that supports bi-encoders since the only change is in the set of document representations to be indexed for NN-search (\reftab{dense_retrieval_pipeline}).
    
    \item 
    \textit{\textbf{Experiments on Public Datasets:}} We evaluate \method as an approach for first-stage retrieval where recall and mean-average precision (MAP) are the key metrics. When applied on top of state-of-the-art (SOTA) single-vector approaches~\cite{DBLP:conf/kdd/NigamSMLDSTGY19,DBLP:conf/emnlp/ReimersG19} in three large diverse datasets, \method consistently delivers $12-28\%$ recall and $4-33\%$ MAP gains (computed over $k=100$ results) by incurring only $1.3\times$ increase in index size. In comparison with the prominent SOTA `multi-vector' \colbert~\cite{DBLP:conf/sigir/KhattabZ20}, \method achieves gains of $3\%-15\%$ recall and $1\%-17\%$ MAP by requiring $4-6\times$ smaller index space. All gains are in absolute percentages, and statistically significant~\cite{DBLP:conf/sigir/YangL99}.

    \item 
    \textit{\textbf{Experiments within E-Commerce Search Engine:}} \method improves a production-quality bi-encoder by $13\%$ recall and $4\%$ MAP (at $k=100$) in terms of offline metrics by increasing the index size by only $1.2\%$ (a large fraction of tail documents were absent from the behavioral graph as is typical in an industrial setting; such documents were not allocated any behavioral representations by \method). In an A/B test, \mvg significantly increased customer engagement metrics without degrading customer-perceived latency.
\end{itemize}

%% file: TEX/2-relatedwork.tex
We review prior dense retrieval (DR) literature, emphasizing the axes which set our work apart via \reftab{relatedwork}. We categorize prior DR methods into two based on whether they learn a single or multiple representation(s) per query and document. 

\begin{table*}[!bt]
    \centering
     \caption{\method vs prior DR approaches on (a) size of nearest-neighbor (NN) index (e.g. HNSW~\cite{DBLP:journals/pami/MalkovY20}), (b) latency or time required for NN search (for the same embedding dimensionality). Let $D$ be the number of documents, $t_d$ ($t_q$) be average number of tokens per document (query). Let $m$ be a fixed hyper-parameter, and $\delta$ be the relative increase in number of document representations due to \method. For conservative values such as $D=1M, t_q=8, t_d=128, m=4, \delta=0.3$, \method achieves  $7-90\%$ less latency while requiring $3-12\times$ smaller index size than other representative multi-vector approaches, e.g. \colbert, \mebert. Thus, \method-based dense retrieval is faster and requires lower memory; as we later show in \refsec{experiments}, \mvg also achieves better retrieval performance.\vspace{-2mm}}
    \label{tab:relatedwork}
    \begin{tabular}{p{5cm}|p{2cm}|p{5cm}|cc}
    \hline
        \multirow{2}{*}{\textbf{Methods}} & \multicolumn{2}{c|}{\textbf{Document representations}} & \multirow{2}{*}{\textbf{Index Size}}& 	\multirow{2}{*}{\textbf{Online Latency}} \\
        & \textbf{Nature} & \textbf{Count} \\ 
    \hline
        Single-vector methods, e.g. \dssm~\cite{DBLP:conf/cikm/HuangHGDAH13}, \textsc{DESM}~\cite{DBLP:journals/corr/MitraNCC16}, \sbert~\cite{DBLP:conf/emnlp/ReimersG19}, ANCE~\cite{DBLP:conf/iclr/XiongXLTLBAO21}& \multirow{2}{*}{semantic only} & \multirow{2}{*}{only one per document} &  \multirow{2}{*}{$\mathcal{O}(D)$} & \multirow{2}{*}{$\mathcal{O}(\log D)$}\\
    %   \multirow{6}{*}{\parbox{1.1cm}{Multiple\\relevance}} 
        \hline 
        \mebert~\cite{DBLP:journals/tacl/LuanETC21}, \textsc{MVR}~\cite{DBLP:conf/acl/ZhangLGJD22}, \textsc{PQR}~\cite{DBLP:conf/acl/TangSJWZW20} & semantic only & multiple, fixed hyperparameter $m$ & $\mathcal{O}(m D)$ & $\mathcal{O}(\log(m D))$\\
        % &  &  & $\mathcal{O}(m D)$ &  $\mathcal{O}(\log(m D))$\\
       \hline
       \colbert~\cite{DBLP:conf/sigir/KhattabZ20}, \textsc{Maize}~\cite{DBLP:journals/naacl/maize}, COIL~\cite{DBLP:conf/naacl/GaoDC21} &  semantic only & multiple, tied to document length $t_d$ & $\mathcal{O}(t_d D)$ & $\mathcal{O}(t_q\log(t_d D))$\\
       \hline
       \method applied on any single-vector approach (this paper) & semantic and behavioral & multiple, variable per document based on hyperparameter $\beta$, $1+\delta$ on average  & \multirow{2}{*}{$\mathcal{O}((1+\delta)D )$} & \multirow{2}{*}{$\mathcal{O}(\log((1+\delta)D))$}\\
       \hline
 \end{tabular}
\end{table*}

\textbf{Single-Vector Bi-encoders} learn a single vector per query and document; such methods are widely used industrially, e.g. Youtube~\cite{DBLP:conf/recsys/CovingtonAS16}, Airbnb~\cite{DBLP:conf/kdd/HaldarRSAZMYTL20}, Amazon~\cite{DBLP:conf/kdd/NigamSMLDSTGY19}.  Early approaches include \dssm~\cite{DBLP:conf/cikm/HuangHGDAH13} and \textsc{DESM}~\cite{DBLP:journals/corr/MitraNCC16} which relied on simple text embedding approaches (e.g. FastText~\cite{DBLP:conf/eacl/GraveMJB17}) and worked well for short-text documents. Recently, the emphasis has shifted to transformer~\cite{DBLP:conf/naacl/DevlinCLT19} based bi-encoders such as \sbert~\cite{DBLP:conf/emnlp/ReimersG19} to better capture semantic similarity for long-text documents by considering the entire context. Several works try to improve negative mining for single-vector bi-encoders, e.g. ADORE~\cite{DBLP:conf/sigir/ZhanM0G0M21}, DPR~\cite{DBLP:conf/emnlp/KarpukhinOMLWEC20} and most recently, ANCE~\cite{DBLP:conf/iclr/XiongXLTLBAO21} which relies on globally hard negatives. Our proposed \method is complementary to any work in this category, and can be applied on top of any single-vector bi-encoder to improve retrieval performance.

\textbf{Multi-Vector Bi-encoders} are becoming increasingly common as a single vector may not suffice to adequately represent a document it~\cite{DBLP:conf/recsys/WestonWY13,DBLP:conf/kdd/PalEZZRL20}. We further subdivide these works in two.
\textbf{\textit{(1)~Fixed count:}} These methods learn $m$ vectors for each document by either directly using the first $m$ BERT token embeddings (ME-BERT~\cite{DBLP:journals/tacl/LuanETC21}, or after clustering all token embeddings into $m$ centroids (PQR~\cite{DBLP:conf/acl/TangSJWZW20}), or by having $m$ \texttt{[CLS]}-like tokens instead of one (MVR \cite{DBLP:conf/acl/ZhangLGJD22}). These methods typically employ a different relevance scoring function during inference than the one used in training. 
\textbf{\textit{(2)~Variable count:}} Here the number of representations learned per document may vary, e.g. based on document length. The most prominent approach here is \colbert~\cite{DBLP:conf/sigir/KhattabZ20} which achieves state-of-the-art performance by preserving all (contextualized) token embeddings from queries and documents and using late-stage interaction in an end-to-end differentiable manner. Several works consider how to optimize memory usage of \colbert while minimally impacting retrieval performance. 
\textsc{Maize}~\cite{DBLP:journals/naacl/maize} uses a residual compression mechanism with a denoised supervision strategy; 
CQ~\cite{DBLP:conf/acl/YangQY22} uses contextual quantization of token embeddings; %by decoupling document specific and document-independent ranking contributions during codebook-based compression. 
COIL~\cite{DBLP:conf/naacl/GaoDC21} leverages contextualized token representations stored in inverted lists; %, bringing together the efficiency of exact match and the representation power of deep language models.
ColBERTer~\cite{DBLP:journals/corr/abs-2203-13088} removes document tokens which minimally impact final scores. 
% These space-improved algorithms tile closely to the structure of \colbert while the proposed \method is flexible to general bi-encoder.% to  
Other methods are \textsc{PRF}~\cite{DBLP:conf/ictir/WangMTO21} which uses pseudo-relevance feedback and MVA~\cite{DBLP:conf/emnlp/ZhouD21} which uses multiple vector attention mechanism. However, a common limitation here is that the learned document representations are all semantic, which is limited by what can be inferred from intrinsic document features (e.g. tokenized text)--especially for short-text documents.
Moreover, the proposed \method is complementary to these DR methods and can be used in combination to create behavioral representations to augment the learned semantic representations.

\textbf{Exploiting graph structure:} Our method uses query-document relevance graph to compute behavioral representations; this is broadly related to the relational learning~\cite{DBLP:conf/sigmod/GetoorM11} and graph neural networks~\cite{DBLP:conf/kdd/LiuTLYZH19,DBLP:conf/kdd/YangPZP0RL20,DBLP:conf/www/FanLL00LLLJY22} but they are typically expensive to serve online at web-scale. The closest work here is simultaneous work~\cite{DBLP:journals/corr/2022graph} which uses graph information within a bi-encoder model. But it does not consider variable number of representations per document.

\textbf{Choice of Baselines:} Given the rapid pace of DR research, and our focus on retrieval settings which are uncommon for DR, a core experimental challenge is select a set of representative baselines which (a) allows us to test all key hypotheses but also (b) is small in number so that we can retrain and tune performance on all three public datasets that we consider. We shortlist two single-vector baselines: classic \dssm and transformer-based \sbert which are better suited for short- and long- text applications respectively. We choose \colbert as the competitive multi-vector baseline. Fortunately, we do not have to consider follow-up works of \colbert which trade-off memory footprint for retrieval performance as we already observe superior performance w.r.t \colbert in Table~\ref{tab:benchmark-public}. 

%% file: TEX/3-prelimnaires.tex
\section{Preliminaries}
\label{sec:prelim}

We provide a brief background on dense retrieval and use this to formally state the problem we tackle in the rest of the paper.

\subsection{Dense Retrieval}
\label{sec:dense-retrieval-background}
Consider historical relevance data $(\queryset, \docset, \weight)$ over a set of queries $\queryset$ and documents $\docset$. For $\query\in\queryset, \doc\in\docset$, the relevance score (e.g. based on past clicks) is noted as $\weight_{\query\doc} \geq 0$. Let $\mathcal{M}$ be a bi-encoder model with query encoder $\mathcal{M}_Q$ and document encoder $\mathcal{M}_D$, where the parameters for the two encoders can be fully or partially shared.
%The goal of first-stage retrieval is to retrieval all relevant documents from document corpus $\docset$ given any user query $q$. 
% There are mainly single- and multiple- relevance computing schemes between documents and queries.

\textbf{Single-Vector Bi-encoders.} Here, we obtain a single representation per query or document. Denote these as $\modelvec{\query} := \mathcal{M}_{Q}(\query)$ and $\modelvec{\doc} := \mathcal{M}_{D}(\doc)$. Typically, $||\modelvec{\query}||_2 = ||\modelvec{\doc}||_2 = 1$, i.e. unit vectors in some $\dimension$-dimensional space, and their relevance score is computed via their dot product or cosine similarity:
\begin{equation}\label{eq:dualencoderrelscore}
\mathtt{rel}\left(\query, \doc\right):=\modelvec{\query}^\top \modelvec{\doc}\\
\end{equation}
As the query and document vectors can be inferred independently, bi-encoders allow us to precompute document vectors and index them in an efficient data-structure, e.g. HNSW~\cite{DBLP:journals/pami/MalkovY20} so that time complexity of nearest-neighbor search to retrieve relevant documents grows only logarithmically in the size of the document corpus.

\textbf{Multi-Vector Bi-encoders.} Here, we obtain one or more representations per query or document. Denote them as $\{\modelvec{\query i}\}_{i\in[1,m_\query]} = \mathcal{M}_{Q}(\query)$ and $\{\modelvec{\doc i}\}_{i\in[1,m_\doc]} = \mathcal{M}_{D}(\doc)$. Many scenarios are possible as shown in \reftab{relatedwork}, e.g. multiple vectors for documents only or for queries only or for both; count of multiple vectors fixed or variable across documents and queries. The most common formulation of multi-vector relevance score~\cite{DBLP:journals/tacl/LuanETC21,DBLP:conf/sigir/KhattabZ20,DBLP:journals/corr/abs-2203-13088,DBLP:conf/acl/YangQY22}, based on \textbf{max-sim} operator over document vectors for a given query vector, is:
\begin{equation}
	\mathtt {rel}(\query, \doc) := \sum_{i\in[1,m_\query]} \max_{j\in[1,m_d]} {\modelvec{\query i}}^\top\modelvec{\doc j}\label{eq:max-sim}
\end{equation}
where again $||\modelvec{\query i}||_2 = ||\modelvec{\doc j}||_2 = 1\ \forall \query, i, \doc, j$. Typically, queries have short text; thus setting $m_\query=1\ \forall\ \query$ helps achieve the best online latency. In this case, retrieval according to \refeq{max-sim} is simply the following: build an NN-index of all representations for all documents and conduct NN-search identical to single-vector case.
% A document $\doc$ is relevant to a query $\query$ if at least one vector $\cidvec{\doc j}$ is close to the query vector $\modelvec{\query}$. This max-sim function  preserves the tractability of nearest neighbor search if we index all the  $m_d$ document vectors and remove  duplicated documents during retrieval.

\subsection{Problem Formulation}
Our goal is to augment semantic document representations learned by a given bi-encoder with behavioral document representations so as to improve the memorization performance of DR approaches. For ease of exposition in the rest of the paper, we assume that bi-encoder under consideration is a single-vector approach; but our formulation and method generalizes to multi-vector bi-encoders in a straightforward manner.
Overall, our problem can be stated as:
\begin{problem}[Behavioral Representation Learning] \label{problem:document-budget}
	\textbf{Given} query-document relevance dataset $(\queryset, \docset, \weight)$, semantic representations of queries and documents $\{\modelvec{\query}\}_{\query\in\queryset}, \{\modelvec{\doc}\}_{\doc\in\docset}$ from a bi-encoder and a budget $M$ on the total number of behavioral representations across documents, \textbf{learn} behavioral document representations $\{\cidvec{\doc j} \mid  j \in[1,\ \numcidvec_\doc]\}_{\doc\in\docset}$ under the given budget so as to \textbf{maximize} the relevance of historically associated query-document pairs.
	\begin{equation} \label{eq:max-relevance}
		\begin{aligned}
			\text{maximize}&\sum_{\query \in \queryset, \doc \in \docset}
			\weight_{\query\doc}\ 
			 \mathtt {rel}(\query, \doc) & \text{s.t.} &\quad \sum_{\doc\in\docset} \numcidvec_\doc \le \budget
		\end{aligned}
	\end{equation}
\end{problem}
Problem~\ref{problem:document-budget} aims to maximize the relevance score of past query-document associations within a given memory budget, essentially seeking better memorization of historical relevance data within the bi-encoder representational space. This sets the stage for an improved recall, which is crucial for DR approaches used for first-stage retrieval. The above maximization objective can be further expanded using the max-sim relevance score from \refeq{max-sim} as
\begin{equation}
    \sum_{\query \in \queryset, \doc \in \docset} \weight_{\query\doc}\ \mathtt {rel}(\query, \doc) = \sum_{\query \in \queryset, \doc \in \docset} \weight_{\query\doc} \max_{j\in[0,m_d]} \modelvec{\query}^\top \cidvec{\doc j}
\end{equation}
where $\cidvec{\doc 0}\ {:}{=}\ \modelvec{\doc}$ is fixed, $||\cidvec{\doc j}||_2 = ||\modelvec{\query}||_2 = 1\ \forall\ \query, \doc, j$, and queries are assumed to represented as single vector each.

% \textbf{Simplication:} 
Jointly solving Problem~\ref{problem:document-budget} for behavioral document representations $\{\cidvec{\doc j}\}_{j \in [1, \numcidvec_\doc], \doc\in\docset}$ and their count $\{\numcidvec_\doc\}_{\doc\in\docset}$ only permits local search algorithms~\cite{Michiels2018} which have no quality guarantees and are also computationally expensive. Therefore, we instead seek a two-step approach which can determine $\{\numcidvec_\doc\}_{\doc\in\docset}$ in a heuristic manner, and then solve the simplified Problem~\ref{problem:document-budget} exactly to compute the optimal behavioral representations $\{\cidvec{\doc j}\}_{j \in [1, \numcidvec_\doc], \doc\in\docset}$.

%% file: TEX/4-generative.tex
\section{Theoretical Connection} \label{sec:generative-model}
We leverage Pitman-Yor Process~\cite{10.1214/aop/1024404422} (PYP) to explain the skewed distribution of query diversity from \reffig{motivation} (right). In this section, we provide background on PYP and draw the connection to document retrieval; later in \refsec{method}, we show how to use it for behavioral budget distribution among documents.

\textbf{Pitman-Yor Process (PYP)}~\cite{10.1214/aop/1024404422} is a discrete-time stochastic process that generalizes the popular Chinese Restaurant Process~\cite{aldous1985exchangeability} to accommodate power-law tails. In a PYP, customers arrive sequentially to be seated in a restaurant with infinite tables, each with infinite capacity. Let $z_n$ be the choice of table for $n^{th}$ customer. After the arrival of $n^{th}$ customer, let $T_{n,k}$ be the number of customers at table $k$ and let $T_n$ denote the number of occupied tables. The table $z_{n+1}$ picked by the $n+1^{th}$ customer is distributed as
\begin{equation}\label{eq:PYP} 
	P(z_{n+1}|z_1,\ldots, z_n)=
	\begin{cases}
		\frac{T_{n,k}-\beta}{n+\alpha}, & 1\le k\le T_n \quad\text{occupied table}\\
		\frac{\alpha+\beta T_n}{n + \alpha}, &  k=T_n+1\quad \text{new table}
	\end{cases}
\end{equation}
where $\alpha > 0$ and $\beta \in (0, 1)$ are concentration parameters: for low values of $\alpha$ and $\beta$, customers tend to crowd around fewer tables. When choosing to sit at one of the occupied tables, observe that customers prefer to sit at popular tables having higher values of $T_{n,k}$; thus PYP naturally captures the `rich get richer' phenomenon.

\textbf{Connection to Document Retrieval:}
We posit that relevant queries for a document have a \textit{clustered distribution}, where each \textit{query cluster} encodes a latent \textit{intent} for accessing the document (e.g. \texttt{kitchen} and \texttt{camping} are two different intents to buy the same \texttt{gas lighter} product). The number of query clusters is not known in advance, and can grow as more queries are observed. There is also a notion of `rich get richer' as the search engine typically uses mechanisms such as query auto-complete or auto-correct to ensure that new queries are similar to common queries in the past. All these characteristics make PYP a natural fit to model queries used to retrieve a document via search engine. Concretely, each document has an associated PYP according to which its associated queries (customers) arrive and map to query clusters or intents (tables).

\textbf{Query Diversity under PYP:} The consequence of the above connection is our ability to leverage classical results from discrete-time stochastic processes to distribute the total budget $\budget$ of behavioral representations among documents in a principled manner. \reflem{pyp} guarantees that the number of tables $T_n$ in a PYP grows sublinearly as $\mathcal{O}(n^\beta)$ in terms of the number of customers $n$.

\begin{lemma}[\citet{pitman2006combinatorial} \textsection 3.3]\label{lem:pyp}
The expected number of tables ($T_n$) occupied by $n$ customers under Pitman-Yor Process (\refeq{PYP}) is:
\begin{align*}
\mathbb{E}(T_n|\alpha,\beta)=\frac{\alpha}{\beta}\left\{\prod_{j=1}^n\frac{\alpha+\beta+j-1}{\alpha+j-1}-1\right\}\asymp\frac{\Gamma(\alpha+1)}{\beta\Gamma(\alpha+\beta)}n^{\beta}
\end{align*}
where $\Gamma$ is the gamma function and $\asymp$ indicates asymptotic equality.
\end{lemma}

Thus, if all documents follow identical and independent Pitman-Yor Processes, the resulting distribution of query diversity across documents (measured in terms of the number of query clusters) also exhibits similar behavior as stated in Corollary~\ref{remark:pyp}. Our proposed approach leverages this result as we discuss next.

\begin{corollary}\label{remark:pyp} For a given document $\doc$ under the Pitman-Yor Process \refeq{PYP}, the expected number of query clusters ($\tilde{\numcidvec_\doc}$) after observing $n_\doc$ queries is: $\mathbb{E}(\tilde{\numcidvec_\doc}|\alpha,\beta)
\asymp\frac{\Gamma(\alpha+1)}{\beta\Gamma(\alpha+\beta)}\left(n_{\doc}\right)^{\beta}$.
\end{corollary}

%% file: TEX/5-method.tex
\section{Our Approach}
\label{sec:method}

\begin{figure}[!t]
	\centering
	\includegraphics[width=0.95\linewidth]{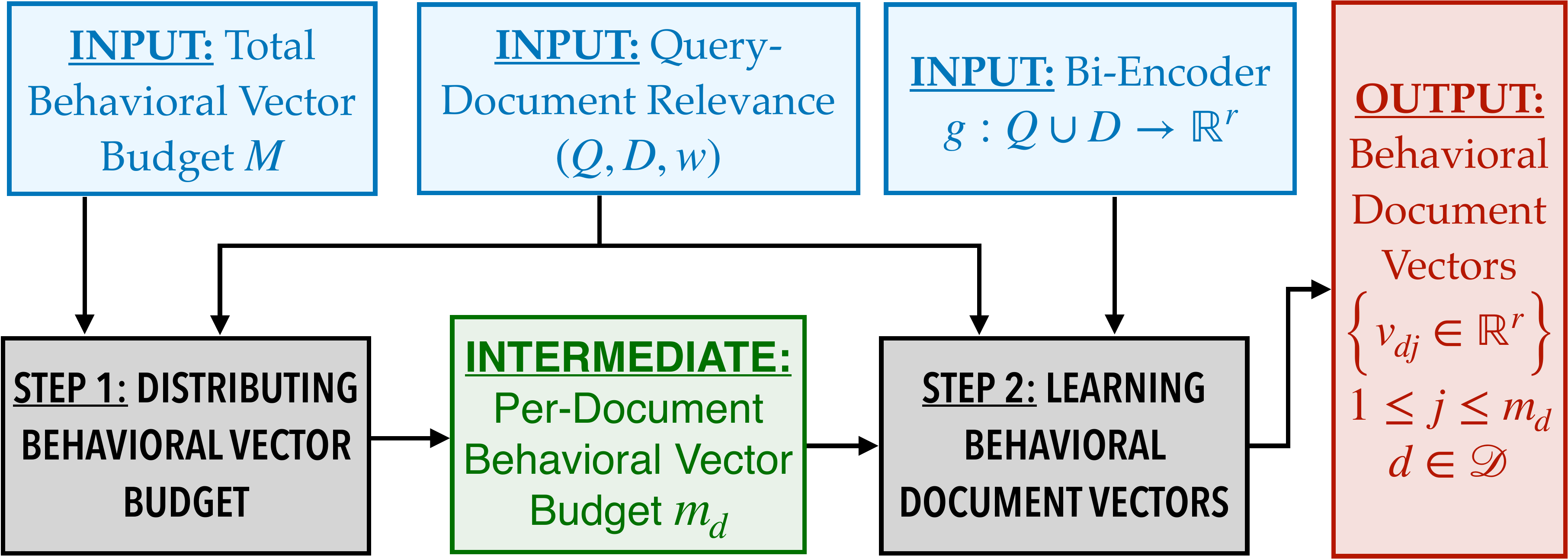}
	\caption{Overview of the proposed method: \method (1) splits a total behavioral vector budget into per-document vector budgets, and (2) uses a Bi-encoder to learn additional document vectors to be indexed for nearest-neighbor search. Both steps utilize the query-document relevance data.\label{fig:method}}
\end{figure}

\input{TEX/algo-EM}

Our proposed approach, called \methodfull (\method), works in two steps as shown in \reffig{method}. First, it determines how to distribute the total budget $\budget$ of behavioral representations across all documents by leveraging the above connection to PYP (\refsec{distributing-budget}). Then, it computes the behavioral representations of a document by clustering its associated queries in the representational space learned by the Bi-encoder (\refsec{learning-vectors}). After describing these steps, \refsec{mvg-practice} reviews how \method fits within the dense retrieval pipeline, and also highlights its desirable properties for use in practice.

\subsection{Step 1: Budget Distribution}
\label{sec:distributing-budget}

How many behavioral representations $\numcidvec_\doc$ out of a total budget $\budget$ should we allocate to a document $\doc$? \method leverages the connection to PYP from \refsec{generative-model} to provide a principled answer: \textit{the number of behavioral representations allocated to a document should be proportional to the number of observed query clusters for the document}. If a document $\doc$ has $n_\doc = \sum_{\query\in\queryset}\mathbf{1}[w_{\query\doc} > 0]$ relevant queries in the training set, then:
\begin{equation}\label{eq:distributing-budget}
 m_d \propto n_\doc^\beta
\end{equation}
where $\beta\in(0,1)$ is now an experimental hyperparameter that allows us to smoothly interpolate between uniform distribution ($\beta=0$) and popularity distribution ($\beta=1$). Due to its ability to model a rich set of budget distribution schemes, the functional form in \refeq{distributing-budget} is widely used in other applications as well~\cite{DBLP:journals/corr/JiVSAD15,DBLP:conf/nips/MikolovSCCD13}.

\subsection{Step 2: Representation Learning}
\label{sec:learning-vectors}
Given the count $\numcidvec_\doc$ of behavioral representations to learn for a document $\doc$, our approach for learning them is to directly optimize the objective in Problem~\ref{problem:document-budget} using max-sim formulation in \refeq{max-sim}. When $\texttt{rel}$ is dot product, it is easy to show \refeq{max-relevance} is the weighted $k$-means clustering objective with $k=\numcidvec_d+1$ and one constrained cluster center $\cidvec{\doc 0}\ {:}{=}\ \modelvec{\doc}$. \refalg{mvg} simply adapts Llyod's algorithm~\cite{DBLP:journals/tit/Lloyd82} for this setting. 

\subsection{\method in Practice}
\label{sec:mvg-practice}
We highlight that \method is a light-weight approach that is easy for dense retrieval practitioners to use. As shown in \reftab{dense_retrieval_pipeline}, \method can be implemented as a single step (step 3) in the standard DR pipeline, with no change in Bi-encoder training or inference procedure (steps 1-2), the  nearest-neighbor (NN) index building step (step 4), or the online retrieval logic (step 5). The only impact of \method is an increase in the number of document representations to be indexed for NN search. But due to efficient NN index data structures, this incurs only a sublinear increase in online retrieval latency~\cite{DBLP:journals/corr/abs-1806-09823}. 

An additional implementation detail when using \method (or any approach learning multiple representations for documents such as ColBERT~\cite{DBLP:conf/sigir/KhattabZ20}) is that the nearest neighbor documents obtained online for a query can contain duplicates as a single document can be retrieved in multiple times due to its many representations. Thus additional deduplication is necessary. However, we find that the deduplication overhead is negligible as online retrieval latency is heavily dominated by that of NN-search.

%% file: TEX/algo-EM.tex
% \caption{Overview of the proposed method: \method (1) splits a total additional vector budget into per-document vector budgets, and (2) uses a dual-encoder to learn additional document vectors to be indexed for nearest-neighbor search. Both steps utilize the query-document relevance data.\label{fig:method}}

\begin{algorithm}[!t]
	\caption{\textbf{\method Algorithm (Step 2) for Learning Behavioral Document Representations}}\label{alg:mvg}
	\begin{algorithmic}[1]
		\Statex \textbf{Input:} Semantic document vector $\modelvec{\doc}$ from bi-encoder; query vectors and relevance weights $\{\modelvec{\query},w_{qd}\}_{\query\in\queryset}$; 
		 budget $\numcidvec_\doc$ of behavioral vectors.
	    \Procedure{\method}{$\modelvec{\doc}, \{\modelvec{\query},w_{qd} \}_{\query\in\queryset},  \numcidvec_\doc$}
	    \State Cluster assignment matrix $S_{qj} {:}{=} 0,\ \forall\ q\in \queryset, j\in[0,\numcidvec_\doc]$
		\For{$q\in\queryset$} {\Comment{random initialization}}
		\State Pick $j'\sim\{0, \ldots, m_d\}$ uniformly at random; 
		\State Set $S_{qj'} := 1$.
		\EndFor
		\State Fix one cluster center $\cidvec{\doc 0} := \modelvec{\doc}$.
		\Repeat 
		\For{$q\in\queryset$ and $j=0, \ldots, \numcidvec_\doc$} {\Comment{update assignments}}
		\State $S_{qj} =\mathbf{1}[ j=\arg\max_{0 \leq j' \leq \numcidvec_\doc} \modelvec{\query}^\top{\cidvec{\doc j'}}]$
		\EndFor
        \For{$j=1 \textbf{ to } \numcidvec_\doc$} {\Comment{update cluster centers}}
		\State $\cidvec{\doc j}= \sum_{q} S_{qj} \weight_{\query\doc} \modelvec{\query}\ /\ \sum_{q}  \weight_{\query\doc} S_{qj}$
		\State $\cidvec{\doc j}=\cidvec{\doc j}/\|\cidvec{\doc j}\|_2$
		\EndFor 
		\Until{matrix $S$ does not change.}\\
		\Return behavioral document representations {$\{\cidvec{\doc j}\}_{j=1}^{\numcidvec_\doc} $.}
	\EndProcedure
	\end{algorithmic}
\end{algorithm}

%% file: TEX/6-experiments.tex
We conduct an extensive evaluation of \method to answer the following questions: \textbf{(Q1)}~Applied on propriety e-commerce product search dataset (which is our motivating problem setting), does \method outperform production dense retrieval baseline in offline metrics? \textbf{(Q2)}~When A/B tested, does \method improve customer engagement within acceptable latency regression? \textbf{(Q3)}~Does \method improve over state-of-the-art dense retrieval approaches across a diverse set of public datasets? \textbf{(Q4)}~What is the behavior of \method with respect to important hyperparameters, e.g. budget for behavioral representations and embedding dimensionality of base bi-encoder? \textbf{(Q5)}~How can we understand the specific examples where \method leads to substantial improvements? \textbf{(Q6)}~How does \method behave on datasets where its motivating factors (e.g. skewed distributions in \reffig{motivation}) are absent? Our findings for these questions are detailed in order in \refsec{prodsearch} through \refsec{negative}.

\textbf{\method Implementation:} We use query-document pairs with positive relevance scores in the training data as input to \method, treating all queries for a document as equally relevant. By default, we use $\beta=0.5$ and learn $\budget_\text{avg}=0.3$ additional vectors per document on average. We apply \method as a wrapper over multiple baseline bi-encoder models described later.

\textbf{Evaluation Metrics:} We use Faiss~\cite{DBLP:journals/tbd/JohnsonDJ21} to retrieve documents that have the highest cosine similarity with the test queries and deduplicate the retrieved document list when evaluating \method. We adopt two of the most commonly used information retrieval metrics, namely, \textit{Recall} and \textit{Average Precision} to evaluate the quality of models in retrieving relevant documents. More specifically, Recall measures the fraction of relevant documents within the retrieved document list, whereas Average Precision measures the goodness of relevant documents by their ranks within the retrieved document list. Due to high variability in the number of relevant documents across queries, we limit the retrieval to top $k \in \{10, 100\}$ documents with the highest cosine similarity. We report Recall@$k$ and MAP@$k$ (mean Average Precision at $k$) as the average of corresponding per-query metrics over all queries, in percentage.

\subsection{Results on Propriety \prodsearch Dataset}
\label{sec:prodsearch}
\input{TEX/6-1-prodsearch}

\subsection{Results on Publicly Available Datasets}
\label{sec:skeweddatasets}
\input{TEX/6-2-public}

\subsection{Effect of Hyper-parameters and Base Model}
\label{sec:hyperparameters}
\input{TEX/6-3-hyperparameters}

\subsection{Case Studies}
\label{sec:casestudies}
\input{TEX/6-4-casestudies}

\subsection{Results on Datasets w/o Skewed Distribution}
\label{sec:negative}
How does \method work on datasets where motivating characteristics from \reffig{motivation} are absent? To answer this, we use the popular MSMarco~\cite{DBLP:conf/nips/NguyenRSGTMD16} question-answering dataset which has flat document popularity distribution, likely due to synthetic curation and heavy preprocessing. \reftab{msmarco} compares the performance of state-of-the-art single-vector (ANCE~\cite{DBLP:conf/iclr/XiongXLTLBAO21}) and multi-vector (ColBERT~\cite{DBLP:conf/sigir/KhattabZ20}) methods to that of \method+ANCE. We see that \method does not bring gains over ANCE as expected; in this case, the improvements from multi-vector ColBERT method are marginal too. Importantly, even when motivating dataset characteristics are absent, \method does not significantly decrease performance compared to underlying bi-encoder.

%% file: TEX/6-1-prodsearch.tex
\begin{table*}[!t]
	\centering
	\caption{Results on proprietary \prodsearch dataset: Metrics are reported over all test queries (``overall''), test queries seen during training (``memorization''), and those not seen during training (``generalization''). Per-query metric values are aggregated by accounting for the traffic contribution of queries. Bold and underline typefaces follow the same convention as in \reftab{benchmark-public}.\vspace{-1mm}}\label{tab:benchmark-prod}
	\scalebox{0.98}{
	\begin{tabular}{r|cc|cc|cc|cc|cc|cc|c}
		\hline
		\multirow{3}{*}{\textbf{Method}}& \multicolumn{6}{c|}{\textbf{Recall@$k$}} & \multicolumn{6}{c|}{\textbf{MAP@$k$}} & \textbf{Index}\\
		&  \multicolumn{2}{c|}{\textbf{Overall}} & \multicolumn{2}{c|}{\textbf{Memorization}} & \multicolumn{2}{c|}{\textbf{Generalization}} & \multicolumn{2}{c|}{\textbf{Overall}} & \multicolumn{2}{c|}{\textbf{Memorization}} & \multicolumn{2}{c|}{\textbf{Generalization}} & \textbf{Size}\\
		& \textbf{$k=10$} &\textbf{$k=100$} & \textbf{$k=10$} &\textbf{$k=100$} & \textbf{$k=10$} &\textbf{$k=100$} & \textbf{$k=10$} &\textbf{$k=100$} & \textbf{$k=10$} &\textbf{$k=100$} & \textbf{$k=10$} &\textbf{$k=100$} & \textbf{(GB)}\\
		\hline
		BM25  & $4.50$ & $13.75$ & $4.13$ & $13.25$ & $13.23$ & $25.47$ & $1.92$ & $2.64$ & $1.72$ & $2.45$ & $6.60$ & $7.13$ & N/A\\
		%Memorization & $\mathbf{40.33}$ & $54.98$& $42.05$ &  $57.31$& $0.00$ & $0.00$ & $32.17$ & $\underline{\mathbf{39.13}}$ & $33.54$ & $\underline{\mathbf{40.79}}$ & $0.00$ &  $0.00$\\
		\prodmodel   & $28.72$ & $55.39$ & $28.50$ &  $55.49$ & $33.86$ &  $53.00$ & $17.13$ & $21.48$ & $17.04$ &  $21.54$ & $19.31$ &  $20.19$ &  $30.0$\\
		\method + \prodmodel & $\mathbf{\underline{33.33}}$ &  $\underline{\mathbf{67.77}}$& $\mathbf{\underline{33.19}}$ &  $\underline{\mathbf{68.14}}$& $\mathbf{\underline{36.76}}$ &  $\underline{\mathbf{59.04}}$ & $\mathbf{\underline{17.55}}$ &  $\mathbf{\underline{25.54}}$ & $\mathbf{\underline{17.47}}$ &  $\mathbf{\underline{25.74}}$ & $\mathbf{\underline{19.51}}$ & $\mathbf{20.73}$ & $30.8$\\
		\hline
	\end{tabular}
	}
\end{table*}

We took a uniform sample from fourteen months of anonymized aggregated search logs from an e-commerce search engine for this experiment. We used data recorded in the first twelve months for training and validation, setting aside the last two months for testing. The document corpus consists of about 60 million products. The baseline dense retrieval model (\prodmodel) follows the network architecture from \cite{DBLP:conf/cikm/HuangHGDAH13}. Both \prodmodel and \method were trained on 1M queries, and tested on 100K queries. For comparison, we also evaluate the standard BM25 model~\cite{DBLP:journals/jasis/RobertsonJ76}, which relies solely on the statistics of overlapping terms between queries and documents.

\renewcommand{\mywidth}{0.5\linewidth}
\begin{figure}[!tb]
	\centering
	\resizebox{\linewidth}{!}{%
		\begin{tabular}{cc}
			\includegraphics[height=\mywidth]{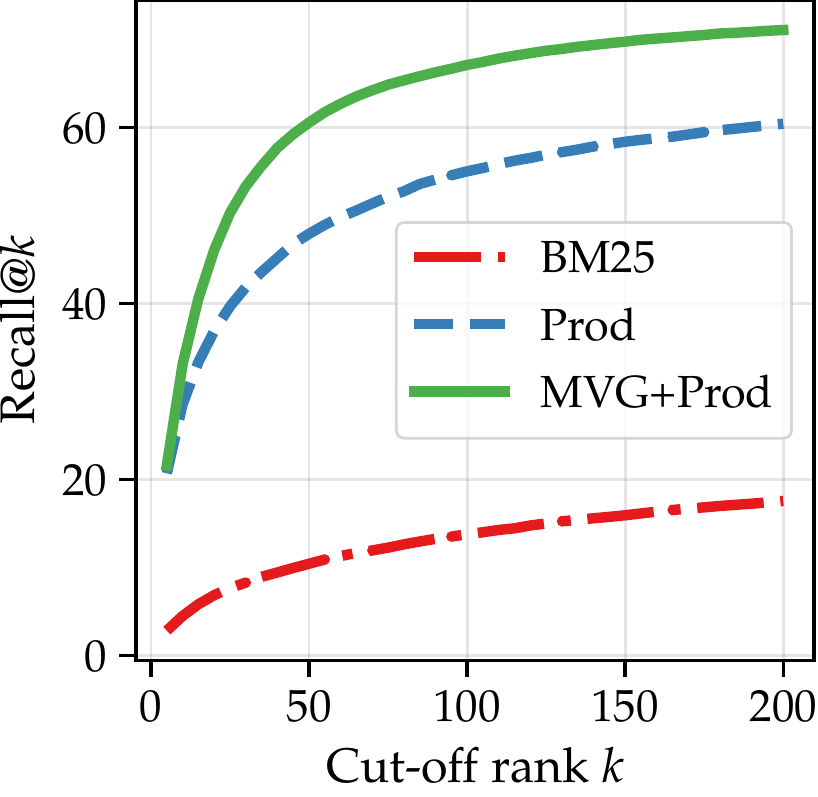} &
			\includegraphics[height=\mywidth]{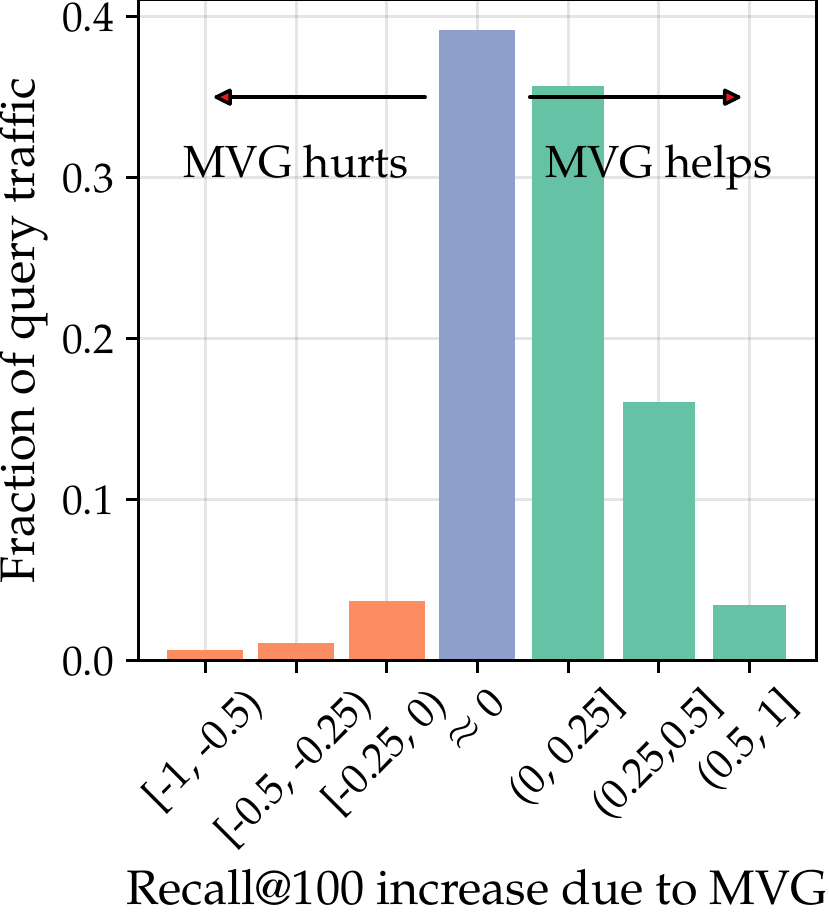}
		\end{tabular}%
	}
	\caption{On ProdSearch Dataset, (Left)~\method improves over \prodmodel on Recall@$k$ for all $k\in[1,200]$. (Right)~\method improves Recall@100 for >$55\%$ of query traffic (green), while hurting only $<5\%$ of query traffic (orange).}\label{fig:recall-prod}
% \vspace{-4mm}
\end{figure}

The \textbf{overall} columns in \reftab{benchmark-prod} show the metrics computed on all test queries. We see that \method applied to \prodmodel (\method+\prodmodel) consistently outperforms \prodmodel alone by improving Recall@$k$ between $4.6\% - 12.4\%$, and MAP@$k$ between $0.4\% - 4\%$. These gains are statistically significant according to two-sided macro-sign test~\cite{DBLP:conf/sigir/YangL99}, and also generalize to $1\leq k \leq 200$ as shown in \reffig{recall-prod} (left). In contrast, BM25 underperforms, indicating its inability to retrieve relevant documents by pure lexical matching from a large corpus. 

The \textbf{memorization} and \textbf{generalization} columns in \reftab{benchmark-prod} show the break-down of metrics by queries that appeared in and were absent from the training set, respectively. Even though \method primarily targets memorization of query-document pairs in the training set, we noticed that it also generalized to new queries. These results showcase the effectiveness of memorization in dense representation space: memorized queries for a document can also help new similar queries discover the relevant document.

\reffig{recall-prod} (right) shows the fraction of query traffic in which \method has gained or lost in Recall@100 compared to \prodmodel alone. We see that the Recall@100 of \method increased in over $55\%$ of query traffic, matched with \prodmodel in about $40\%$ of query traffic, but lost only in less than $5\%$ query traffic. This analysis shows that \method is effective and safe to use even though the additional queries may also, in theory, invite more irrelevant documents into retrieval results. 

%\reffig{recall-prod} (right) shows \method boosts the relevance scores positive query-document pairs much more than negative ones, \method is able to rectify these memorization gaps.

\begin{table}[!t]
    \centering
    \caption{A/B test results showing \% improvements from \method over production system. Underline indicates statistical significance while arrows indicate whether higher ($\uparrow$) or lower ($\downarrow$) is better on a metric. All metrics are defined in \refsec{abtest}.\vspace{-2mm}}
    \label{tab:abtest}
    \begin{tabular}{ccccc}
        \toprule
         \textbf{Units ($\uparrow$)} & \textbf{PS ($\uparrow$)} & \textbf{E+S ($\uparrow$)} & \textbf{SR ($\downarrow$)} & \textbf{Latency P90 ($\downarrow$)} \\
         \midrule
         \underline{+$0.24\%$} & +$0.06\%$ & \underline{$-0.18\%$} & \underline{-$1.21\%$} & $-3$ms\\
         \toprule
    \end{tabular}
    \vspace{-2mm}
\end{table}

\subsection{A/B Testing in E-Commerce Product Search}
\label{sec:abtest}
An e-commerce service typically collects first-stage retrieval results from several sources like lexical matchers (inverted indices), semantic matchers (dense retrieval methods), upstream machine learning models, and advertised products. To evaluate \method online, we replace only the production dense retrieval method (X) with its \method-augmented version (X+\method) and measure improvements in \textbf{(a) customer engagement:} number of products purchased \textbf{\textit{(Units)}} and the amount of product sales \textbf{\textit{(PS)}}, \textbf{(b) relevance quality:} percentage of exact or substitute matches \textbf{\textit{(E+S)}} within the top-16 products, \textbf{(c) sparse results:} percentage of queries with less than 16 products retrieved \textbf{\textit{(SR)}}, and \textbf{(d) latency P90:} value of latency at 90th percentile traffic. \reftab{abtest} tabulates the results. We observe that \method brings statistically significant improvements in customer engagement metrics and sparse results without impacting the overall latency. However, there is a slight degradation in relevance quality metrics, which we discuss in \refsec{conclusion} and leave for future work.

%% file: TEX/6-2-public.tex
We evaluate \method on publicly available datasets which have the characteristics that motivated our approach: skewed distribution in frequency and diversity of queries across documents from \reffig{motivation}.

\textbf{Dataset Description:} 
\textbf{(1)~\eurlex~\cite{DBLP:journals/corr/abs-1905-10892}}\footnote{https://huggingface.co/datasets/eurlex} for labeling legal documents with one or more EUROVOC concepts (e.g. international affairs, data processing). We use document titles as queries and concepts as documents for dense retrieval. The dataset consists of 4271 documents and 45K, 6K, and 6K queries in the training, validation, and test sets, respectively.
\textbf{(2)~\pubmed\footnote{\url{https://www.kaggle.com/bonhart/pubmed-abstracts}}}~\cite{10.1093/comjnl/bxaa073} for classification of biomedical articles into one or more NIH Medical Subject Headings (MeSH) (e.g. fibrosis, humans, and mutation). We use article titles as queries, and MeSH labels as documents. After preprocessing and splitting, the dataset consists of 27K documents and 1.2M, 1K, and 100K queries for the training, validation, and test sets, respectively. 
\textbf{(3)~WikiSeeAlso\footnote{\url{http://manikvarma.org/downloads/XC/XMLRepository.html}}}~\cite{DBLP:conf/mike/LabhishettySNC17} for predicting the titles of related Wiki pages from a given webpage title. The dataset consists of 300K documents and 500K, 138K, and 177K queries in the training, validation, and test sets, respectively.

\begin{table*}[!t]
	\centering 
	\caption{\method outperforms baselines on publicly available datasets in Recall@$k$ and MAP@$k$: Bold indicates best method under each metric within each row group, while underline signifies statistically significant differences ($p < 0.001$) with respect to the second best method in the same row group according to a two-sided macro-sign test~\cite{DBLP:conf/sigir/YangL99}.\vspace{-3mm}}\label{tab:benchmark-public}
	\begin{tabular}{r|ccc|ccc|ccc}
		\hline
		\multirow{3}{*}{\textbf{Method}} & \multicolumn{3}{c|}{\textbf{Recall@$100$}} & \multicolumn{3}{c|}{\textbf{MAP@$100$}} &\multicolumn{3}{c}{\textbf{Index Size (\# Floats in Millions)}} \\
		& \textbf{\eurlex} &\textbf{\pubmed} &\textbf{\wikiseealso} &  \textbf{\eurlex} &  \textbf{\pubmed} & \textbf{\wikiseealso} &\textbf{\eurlex} &  \textbf{\pubmed} & \textbf{\wikiseealso} \\
		\hline
		BM25  & $18.95$  & $20.55$  & $25.90$  &  $7.26$  & $11.22$ & $9.55$  & \multicolumn{3}{c}{\multirow{2}{*}{N/A}}\\
		Popularity& $46.26$	 & $33.65$  &	$0.20$  & $8.35$  & $10.35$ & $0.12$   \\
		\hline
		
		\dssm~\cite{DBLP:conf/cikm/HuangHGDAH13}  & $67.20$  & $20.24$  & $35.44$  &  $31.83$  & $6.58$  & $11.42$ & $1.84$   & $7.52$ & $78.95$ \\
		\method + \dssm  &  $\underline{\mathbf{81.62}}$ &  $\underline{\mathbf{49.01}}$ & $\underline{\mathbf{37.90}}$    & $\underline{\mathbf{51.28}}$ & $\underline{\mathbf{16.70}}$ &  $\underline{\mathbf{13.73}}$  & $2.40$ & $9.77$  & $102.63$\\
		\hline

		\sbert~\cite{DBLP:conf/emnlp/ReimersG19}  & $50.58$  & $35.49$ & $29.38$ &  $14.75$  & $6.87$ & $9.66$  & $5.53$ & $22.56$  & $236.86$\\
		\method + \sbert & $\underline{\mathbf{78.15}}$  & $\underline{\mathbf{59.62}}$ & $\underline{\mathbf{42.12}}$ & $\underline{\mathbf{47.87}}$ &  $\underline{\mathbf{17.73}}$ & $\underline{\mathbf{13.78}}$  & $7.19$ & $29.32$  & $307.92$\\
		\colbert~\cite{DBLP:conf/sigir/KhattabZ20}  &$73.47$ & $44.88$ & $39.01$  & $31.24$ & $14.35$ & $12.70$ & $30.66$ & $170.22$ & $1748.16$\\
		\hline
	\end{tabular}
\end{table*}

\textbf{Baselines:} We compare \method to the following:
(i) \textbf{BM25}~\cite{DBLP:journals/jasis/RobertsonJ76} as before.
(ii) \textbf{Popularity}, which ranks documents based on the count of unique queries associated in training set. Even though the same document list is applied to all test queries, this baseline remains competitive on datasets with skewed distribution of association between queries and documents.
(iii) \textbf{Deep Structured Semantic Model} (\dssm)~\cite{DBLP:conf/cikm/HuangHGDAH13}
%computes the query and document embeddings by averaging their token embeddings followed by a non-linear activation function. The token embeddings are typically shared between the query and document encoders. We use
using SentencePiece Byte-Pair Encoding~\cite{DBLP:conf/emnlp/BostromD20} to tokenize text. The tokenizers are trained on both queries and documents to learn a vocabulary of size 40K for \eurlex, 300K for \pubmed, and 100K for \wikiseealso. The vocabulary sizes were set to roughly half the number of unique words in the respective datasets.
(iv) \textbf{SentenceBERT} (\sbert)~\cite{DBLP:conf/emnlp/ReimersG19} %, DBLP:conf/emnlp/GaoYC21} \dhivyacomment{Nan, why do we have two citations for SBert?}
%uses the state-of-the-art Siamese BERT architecture. Query and document vectors are obtained from their \texttt{[CLS]} token embeddings, which is fine-tuned by \texttt{CosineSimilarityLoss} on the query-document relevance data. We 
initialized using domain-specific pretrained BERT: \texttt{LegalBERT}\footnote{\url{https://huggingface.co/nlpaueb/legal-bert-base-uncased}} for \eurlex, \texttt{PubMedBERT}\footnote{\url{huggingface.co/microsoft/BiomedNLP-PubMedBERT-base-uncased-abstract}} for \pubmed, and \texttt{BERTBase}\footnote{\url{https://huggingface.co/bert-base-uncased}} for \wikiseealso and fine-tuned using default hyper-parameter settings.  (v) \textbf{ColBERT}~\cite{DBLP:conf/sigir/KhattabZ20} using the default configuration. This is the state-of-the-art dense retrieval method learning multiple representations per document. %\footnote{\url{https://github.com/stanford-futuredata/ColBERT}}

\textbf{\method vs Single-Vector Bi-encoders:} We apply \method as a wrapper on top of \dssm and \sbert. We do not compare with solutions built for the datasets that are not of bi-encoder architecture because they are not compatible with \method. 
\reftab{benchmark-public} summarizes the performance of all approaches on all the public datasets. 
Popularity is more competitive than BM25 on \eurlex and \pubmed which exhibit skewed query-document distribution but not on \wikiseealso that has the least skewed distribution among the three datasets as shown in Figure~\ref{fig:motivation}.
%\dhivyacomment{Nan: Explain why popularity for wikiseealso is low. otherwise people will think it's a bug} 
The dense retrieval approaches \dssm and \sbert achieve higher metrics, with \sbert yielding higher metrics on \pubmed dataset where semantic understanding of documents is more crucial. Importantly, \method consistently outperforms both these baselines, delivering upto $27.27\%$ gain in Recall@100 and $22.95\%$ gain in MAP@100 -- at the cost of only $1.3\times$ increase in index size. Thus \method can bring value in a wide variety of settings across diverse datasets and underlying bi-encoders. 

\textbf{\method vs Multi-Vector Bi-Encoder:} Importantly, \method also improves over the state-of-the-art multi-vector \colbert baseline which learns multiple semantic (but no behavioral) representations per document (and query). Both \method and \colbert can be viewed as improvements over the same single-vector \sbert bi-encoder to incorporate multiple behavioral representations and multiple semantic representations respectively. However, as seen in
\reftab{benchmark-public}, on datasets with motivating skewed distributions, behavioral representations turn out to be much more effective. Thus, \method+\sbert outperforms \colbert by a statistically significant $3-15\%$ recall and $1-17\%$ MAP by requiring $4-6\times$ smaller space as measured by number of million floating points stored within the NN-index. Our results suggest that improvements from \method will likely persist over recent dense retrieval methods which optimize the memory overhead of \colbert by sacrificing some retrieval performance~\citet{DBLP:conf/naacl/GaoDC21}. Finally, \reffig{extra_space_comp} illustrates that wide spectrum of memory overhead that \method offers a practitioner to play with, compared to \colbert which is a fixed point in the figure; this makes \method better suited to practical settings with stricter latency and memory requirements.

%% file: TEX/6-3-hyperparameters.tex
We study how the performance depends on choice of hyperparameters for \method and the capacity and quality of base bi-encoder.

\textbf{Budget Hyper-parameters:} 
We vary total budget $\budget$ by varying the average number of behavioral vectors per document $\budget_{\text{avg}}\in[0.01,3]$. 
\reffig{dim} (left) illustrates how the relative budget allocated to head, torso, tail documents varies with $\beta$ (these are the top, middle, bottom 33$\%^{le}$ documents based on number of queries $n_\doc$). As $\beta\rightarrow0$, all documents get equal budget, and when $\beta\rightarrow1$, $80\%$ budget is allocated to head documents which account for most unique queries. Intermediate values provide more reasonable budget distributions, and thus we use them to chart the variation of Recall@100 with $\budget_{\text{avg}}$ in \reffig{dim} (right). For a given $\beta$, as we increase $\budget_\text{avg}$, the recall steeply improves due to the improved memorization. 
The best recall is achieved for $\beta=0.3,\budget_{\text{avg}}=0.5$. The recall falls gradually as $\budget_{\text{avg}}$ is increased beyond its optimal value however, suggesting the \method overfits to training data and hurts generalization. Larger values of $\beta$ overfit more severely to the head documents, and hence show a steeper decrease in this regime. Importantly, observe that for \textit{all} hyper-parameters, \method consistently matches or outperforms the \prodmodel baseline (gray dashed line).

\renewcommand{\mywidth}{0.2\linewidth}
\begin{figure}[!tb]
	\centering
	\resizebox{1\columnwidth}{!}{%
		\begin{tabular}{cccc}
			\includegraphics[height=\mywidth]{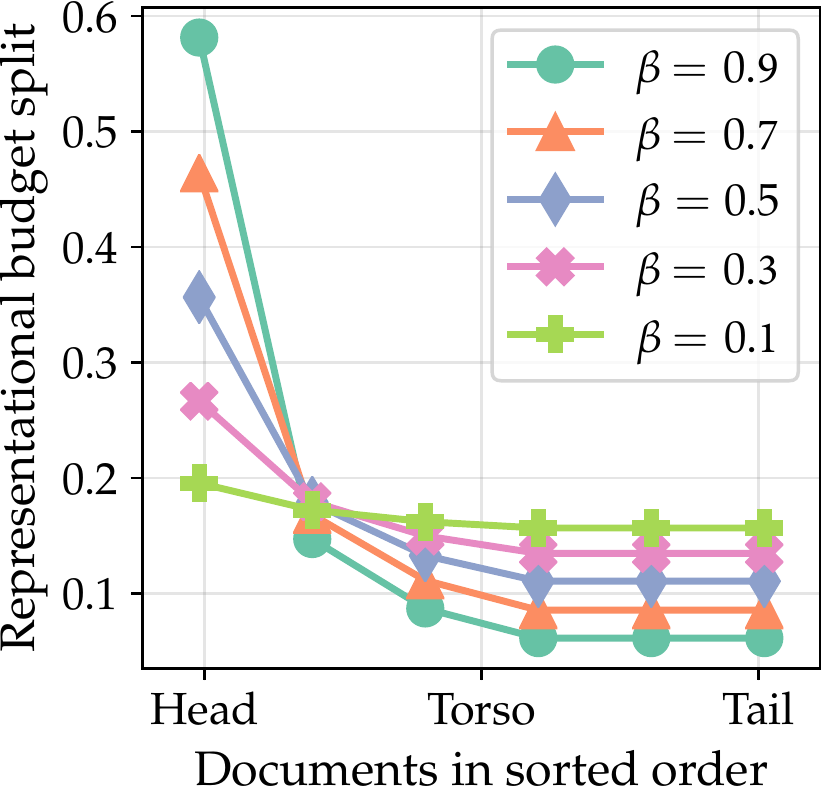} &
			\includegraphics[height=\mywidth]{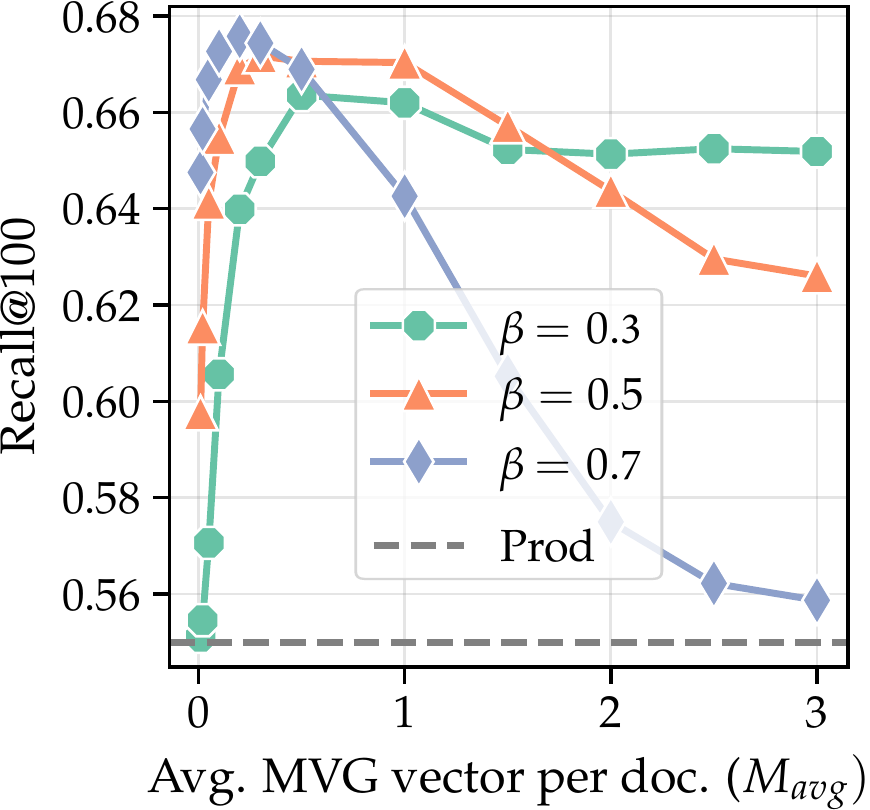} 
		\end{tabular}%
	}
	\caption{On \prodsearch dataset, (left)~distribution of representational budget among documents based on $\beta$.  (right)~\method outperforms \prodmodel for all hyper-parameter settings.}\label{fig:dim}
\end{figure}

\renewcommand{\mywidth}{0.48\columnwidth}
\begin{figure}[!t]
	\centering
	\resizebox{1\columnwidth}{!}{%
		\begin{tabular}{ccc}
			\includegraphics[height=\mywidth]{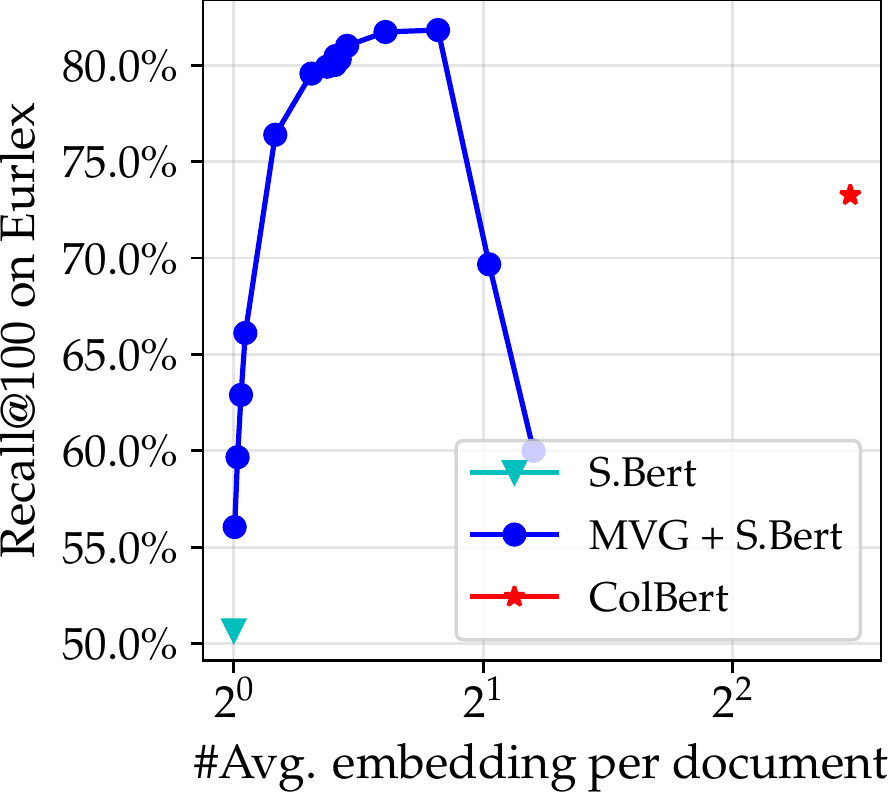}&
		\includegraphics[height=\mywidth]{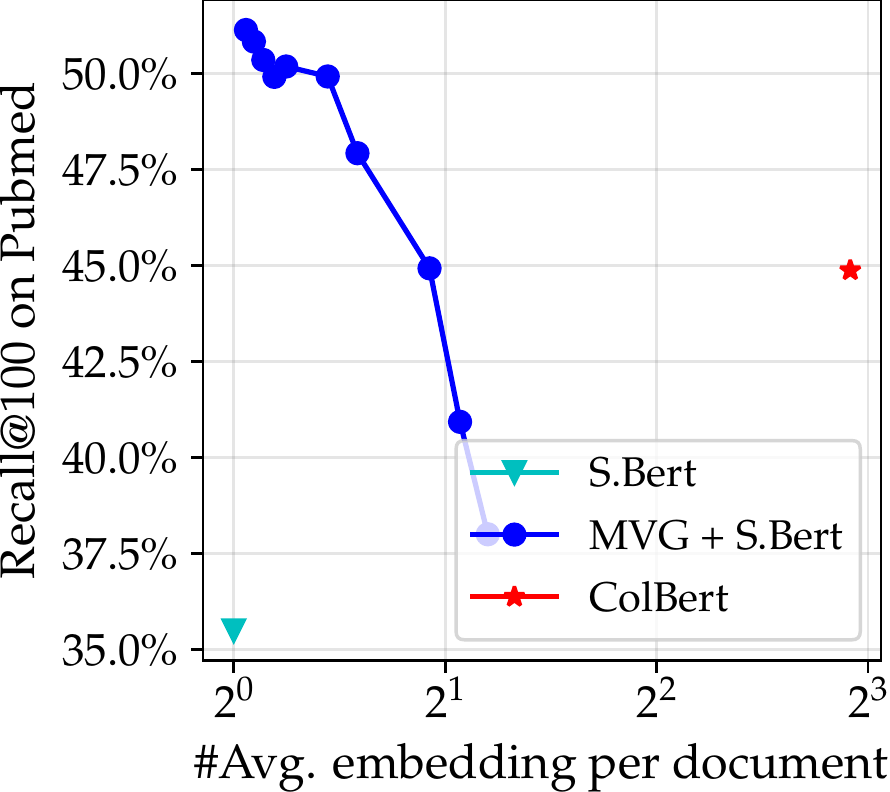}
		\end{tabular}%
	}
	\caption{For a given bi-encoder (here, \sbert), \colbert incurs a fixed memory overhead which depends on document length but is out of practitioner's control. In contrast, \method+\sbert exposes a wider spectrum of memory overhead that is better suited to practical settings having strict memory or latency limits.} 
	\label{fig:extra_space_comp}
\end{figure}

\textbf{Embedding Dimension of Base Bi-Encoder:}  In Figure~\ref{fig:hypers} (left), we vary the embedding dimensionality $\dimension$ of \dssm, and compare it to $\dimension$-dimensional \dssm+\method with $1.3\times$ memory overhead. We see that the recall of \dssm increases steadily with larger dimensions. This behavior is expected based on~\cite{DBLP:conf/acl/0001G20}; and \method+\dssm follows this pattern too. In particular,  
at $\dimension=32$, \method brings a $18\%$ recall improvement (from $58\%$ to $76\%$) compared to \dssm with just $1.3\times$ memory overhead. However, even with $100\%$ memory overhead (doubling dimensionality), \dssm with $\dimension=64$ attains a recall of just $63\%$. This shows that index size increase by allocating non-uniform representational budget (e.g. more vectors to head documents) performs better than uniformly increasing representational budget across all documents (e.g. dimensionality increase).

\renewcommand{\mywidth}{0.45\linewidth}
\begin{figure}[!t]
	\centering
	\resizebox{\columnwidth}{!}{%
		\begin{tabular}{cc}
			\includegraphics[height=\mywidth]{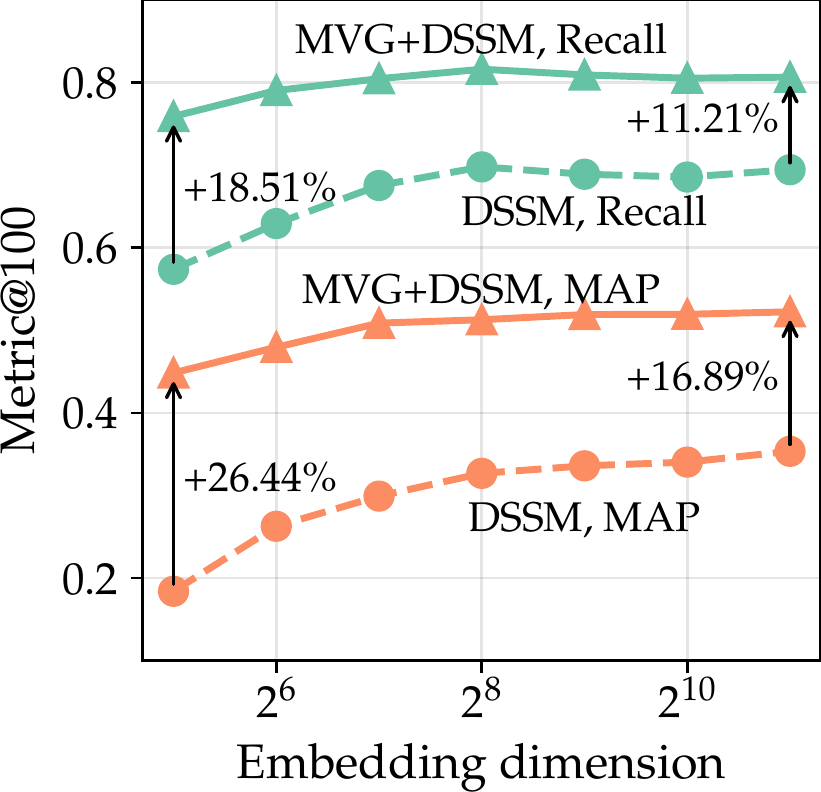} &
			\includegraphics[width=\mywidth,height=\mywidth]{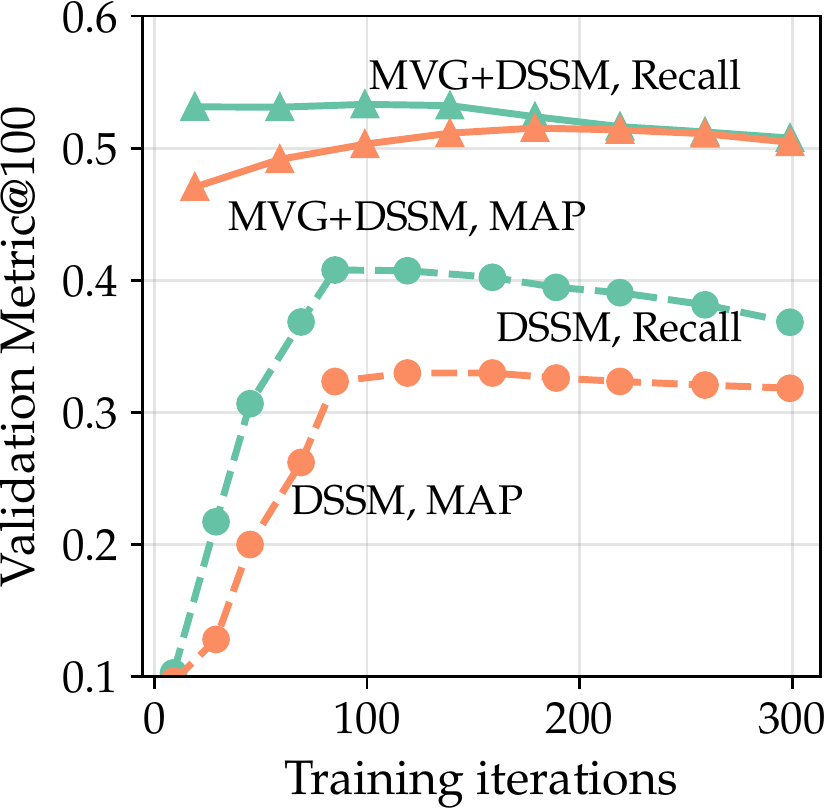} \\
		\end{tabular}%
	}
	\caption{On Eulex dataset, (Left)~\method improvement with variation of embedding dimensions (Right)~\method improvement throughout the training iterations.} \label{fig:hypers}
\end{figure}

\textbf{Base Model Quality:} To effectively produce base model \dssm of varying quality, we set a limit on the training iterations (or training set size) for base model. Figure~\ref{fig:hypers}(right) shows the metrics on \eurlex dataset for base model \dssm and \method (applied to \dssm) at various training iterations. The Recall@100 and MAP@100 of \dssm model gradually increase as training progresses before showing sign of model overfitting towards the end where metrics decreased. In the early stage of training, \method improves over \dssm with $42.8\%$ for Recall@100 and $37.4\%$ on MAP@100, which indicates that \method can improve on \dssm even when it has not fully converged. At the stage where \dssm attains the highest Recall@100, \method shows $12.5\%$ and $19.2\%$ improvement over \dssm on Recall@100 and MAP@100 respectively.
We also evaluate the impact of \method over pretrained BERT models with no task-specific finetuning on \eurlex dataset: \texttt{BERTBase} pretrained on general data source, \texttt{LegalBERT} pretrained on European legislation documents. As shown in Table~\ref{tab:bad-base}, \method brings large recall and MAP improvements, suggesting that applying \method over pretrained bi-encoders can potentially serve as a light-weight alternative to expensive fine-tuning when memorization is the primary objective.

\begin{table}
\centering
\caption{On Eurlex dataset, \method improves over \textsc{BertBase} and \textsc{LegalBert} models on two metrics.} \label{tab:bad-base}
\begin{tabular}{r|cc|cc}
\hline
\multirow{3}{*}{\textbf{Method}}&  \multicolumn{2}{c|}{\textbf{Recall@$k$}} & \multicolumn{2}{c}{\textbf{MAP@$k$}} \\
& \textbf{$k=10$} &\textbf{$k=100$} & \textbf{$k=10$} &\textbf{$k=100$}  \\
\hline
\BertBase &  $0.66$ & $3.42$ & $0.23$ & $0.31$ \\
\method + \BertBase & $55.84$ & $76.20$ &$43.46$ & $46.58$\\
\hline
\LegalBert & $3.74$ & $11.53$ &$1.92$ & $2.26$  \\
\method + \LegalBert & $56.15$ & $77.15$ & $43.53$ & $46.90$\\
\hline
\end{tabular}
\end{table}

%% file: TEX/6-4-casestudies.tex
\renewcommand{\mywidth}{0.45\linewidth}
\begin{figure}[!tb]
	\centering
	\resizebox{\linewidth}{!}{%
		\begin{tabular}{cc}
			\includegraphics[height=\mywidth]{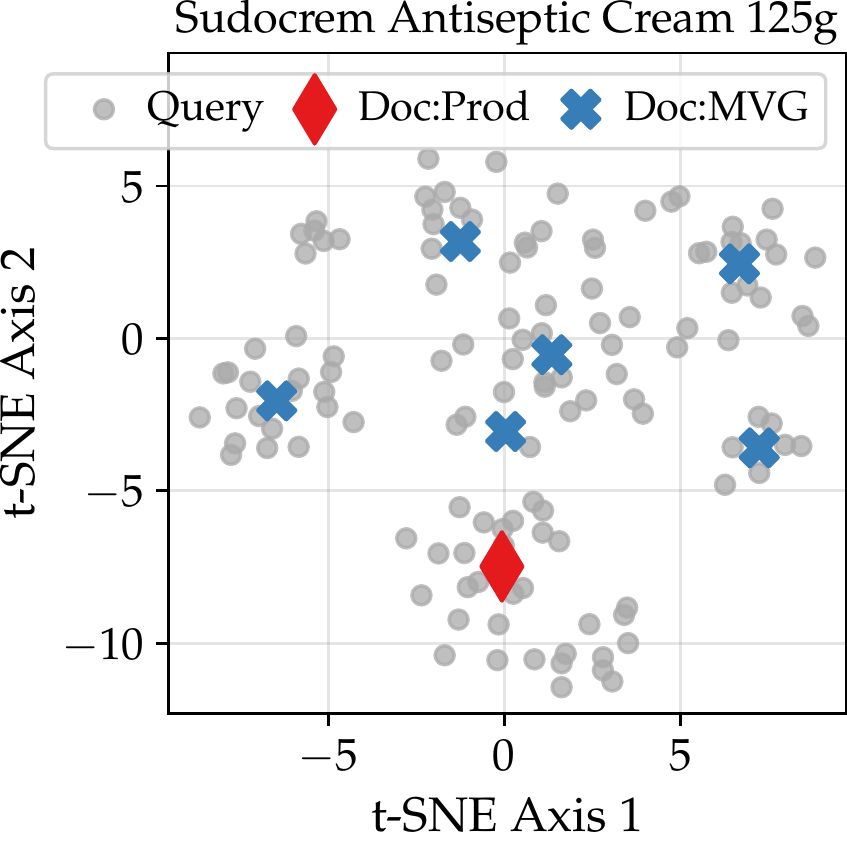} &
			\includegraphics[height=\mywidth]{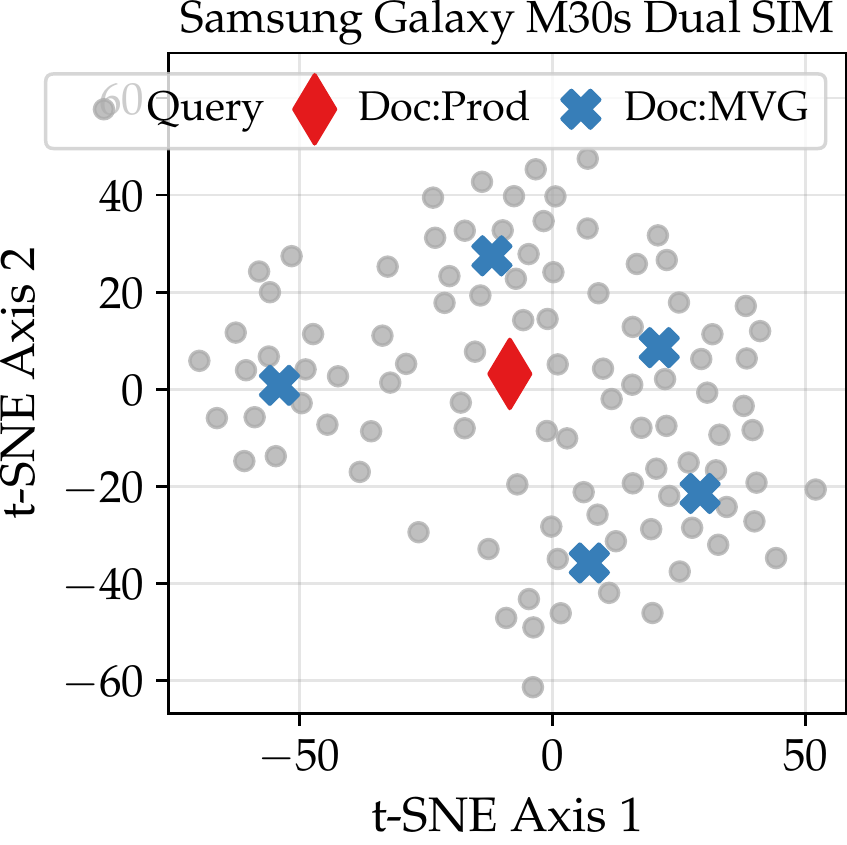} \\
		\end{tabular}%
	}
	\caption{On \prodsearch dataset,  t-SNE~\cite{JMLR:v9:vandermaaten08a} visualization of vectors query vectors (grey dots), document vectors (red diamond). \method captures the relevance of far-away queries by having several auxilary vectors (blue cross).}\label{fig:casestudies}
\end{figure}

\renewcommand{\mywidth}{0.55cm}
\begin{figure*}[!t]
	\begin{minipage}{\textwidth}
		\begin{minipage}[c]{0.25\textwidth}
			\centering
			\includegraphics[width=0.9\linewidth]{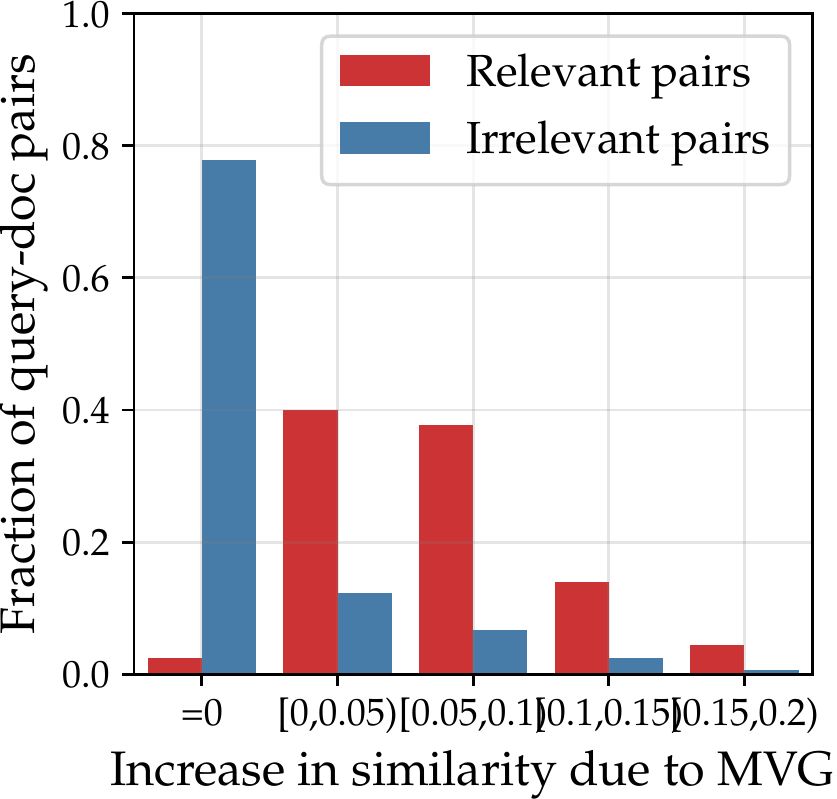}
			\captionof{figure}{\method selectively boosts similarity scores of relevant query-document pairs.\label{fig:cosinediffhist}}
		\end{minipage}
		\hfill
		\begin{minipage}[c]{0.7\textwidth}
			\centering
			\captionof{table}{Examples from \prodsearch training dataset demonstrating gaps in memorization of Prod model: the similarity scores are low, leading to ranks worse than 100. \method rectifies these gaps, leading to improved similarity scores and ranks.\label{tab:motivation}}
			\begin{tabular}[c]{rr|cccc}
				\toprule
				\multirow{2}{*}{\textbf{Query}}&  \multirow{2}{*}{\textbf{Document (Product)}} & \multicolumn{2}{c}{\textbf{Prod}} & \multicolumn{2}{c}{\textbf{\method+ Prod}}\\
				& & \textbf{Sim} & \textbf{Rank} & \textbf{Sim} & \textbf{Rank} \\
				\midrule
				{ automatic water plant} & { DIY Automatic Drip Irrigation Kit} & 0.65 & >500 & 0.72 & 74 \\
				art for toddlers & { Doodle Dog Arts and Crafts} & 0.63 & 265 & 0.75 & 24 \\
				{ gym glives} & { Adidas Performance Gloves} & 0.64 & >500 & 0.73 & 81 \\
				{ label manager} & { Brother Easy Portable Label Maker} & 0.66 & 112 & 0.75 & 8\\
				{ arm mobile holder} & { Adidas Sports Armband for iPhone X} & 0.70 & >500 & 0.80 & 19 \\ % 
				\bottomrule
			\end{tabular}
		\end{minipage}
	\end{minipage}
\end{figure*}

\reffig{casestudies} depicts case studies from \prodsearch dataset to help understand where and how \method helps compared to the underlying bi-encoder.
For each document, we display its semantic vector learned by \prodmodel as a red diamond, behavioral vectors learned by \method as blue crosses, and queries associated with the document in the training data as grey dots--after projecting all vectors into two dimensions using t-SNE~\cite{JMLR:v9:vandermaaten08a}. While $L_2$ distances in the projected space only approximately preserve cosine similarities in the original space, these plots provide an insight into the relative distances among all vectors of interest in the high-dimensional space. After examining such plots for many documents, we discover two types of scenarios where \method proves most useful over \prodmodel bi-encoder.

\reffig{casestudies} (left) depicts the first success scenario: the \prodmodel's semantic document vector captures only a single group of relevant queries (red diamond is near one grey cluster only), and \method learns behavioral vectors (blue crosses) to capture other groups of relevant queries. Observe that the blue crosses are not only situated close to tight grey clusters (likely queries sharing the same intent) but are also far away from the red diamond (semantic document vector). This visually confirms that \method learns behavioral vectors which complement the semantic \prodmodel vector of the document, and thus representation power is not wasted trying to capture the same set of queries more than once.
\reffig{casestudies} (right) shows the second success scenario: the \prodmodel document vector is unable to capture any single group of relevant queries sufficiently well (red diamond is in a sparse central region away from most grey points). In this case, the \prodmodel document vector is essentially wasted, while \method successfully learns multiple behavioral vectors which better represent the dense clouds of grey points.

\begin{table}[!tb]
    \centering
    \caption{On MSMarco dataset~\cite{DBLP:conf/nips/NguyenRSGTMD16} with flat document popularity distribution, \method does not improve over single-vector or multi-vector baselines; notably, it does not hurt either.}
    \label{tab:msmarco}
    \begin{tabular}{l|ccc}
    \hline
       \textbf{Recall}@$k$ &	\textbf{ANCE}~\cite{DBLP:conf/iclr/XiongXLTLBAO21} &	\textbf{ANCE+MVG} & \textbf{ColBERT~\cite{DBLP:conf/sigir/KhattabZ20}}\\ \hline
$k=10$ &	$58.35$ &	$58.35$ &  $60.83$ \\
$k=100$ &	$85.24$ &	$85.24$ &  $85.82$\\
% $k=500$ &	$93.72$ &	$93.72$ &  $93.73$ \\
$k=1000$ &	$95.87$ &	$95.66$ &  $96.18$ \\
\hline
    \end{tabular}
\end{table}

\reftab{motivation} provides examples of query-document pairs from \prodsearch training dataset that were missed by the top-100 retrieved documents of the \prodmodel model, but were correctly captured by \method. 
For these cases, a single vector for a document was insufficient; as \method boosts the relevance scores of positive query-document pairs much more than negative ones (Figure~\ref{fig:cosinediffhist}), \method is able to rectify these memorization gaps.

%% file: TEX/8-conclusion.tex
We presented a simple yet effective approach called \method to improve any given bi-encoder model for first-stage retrieval by incurring only a marginal memory overhead. \method was motivated by the skewed distributions in document popularity and query diversity in real-world retrieval settings, which have thus far largely been overlooked in the dense retrieval literature. To correct for this, our main ideas were to (1) carefully consider the distribution of representational budget among documents based on document properties (e.g. popularity); (2) learn behavioral representations for documents which can more flexibly memorize past query-document associations compared to the typical semantic representations which are limited by what can be inferred from intrinsic document features (e.g. tokenized text that is short).  Extensive experiments over three large public datasets, a proprietary e-commerce search dataset, and an A/B test consistently demonstrated the merits of our ideas and the specific way in which \method incorporates them--as long as the dataset characteristics align with our motivations. When dataset characteristics deviate, however, (e.g. MS Marco) we saw that \method does not bring gains; but it does not significantly hurt either. 

We hope that our work paves the way for next generation of dense retrieval approaches which more carefully consider the dual questions of representational budget distribution, and of jointly learning semantic and behavioral representations. Some concrete directions for future work are: (a) improving robustness to noise and overfitting (\reftab{abtest}, \reffig{extra_space_comp}) by using a discriminative approach instead of clustering, (b) hyperparameter-free automatic way to infer the number of behavioral representations per document.